\documentclass[11pt]{article}

\usepackage[final]{acl}

\usepackage{times}
\usepackage{latexsym}

\usepackage[T1]{fontenc}

\usepackage[utf8]{inputenc}

\usepackage{microtype}
\usepackage{inconsolata}

\usepackage{graphicx}

\usepackage{xcolor}
\usepackage{subcaption}

\usepackage{color, colortbl, xcolor}
\usepackage{url}
\usepackage{subcaption}
\usepackage{textcomp}
\usepackage{soul}
\usepackage{multirow}
\usepackage{enumitem}
\usepackage{mathtools}
\usepackage{siunitx}
\usepackage{tabularx}

\usepackage{bbding}
\usepackage{pifont}
\usepackage{wasysym}
\usepackage{amssymb}
\usepackage{booktabs} 
\usepackage{longtable,booktabs}

\usepackage{multicol}
\usepackage{natbib}  

\usepackage{arydshln}
\definecolor{linkColor}{RGB}{6,125,233}
\definecolor{green}{rgb}{0.0, 0.65, 0.31}
\definecolor{bleudefrance}{rgb}{0.19, 0.55, 0.91}
\definecolor{ceruleanblue}{rgb}{0.16, 0.32, 0.75}
\definecolor{grey}{HTML}{969696}
\definecolor{violet}{HTML}{756bb1}
\definecolor{dgrey}{HTML}{01665e}
\definecolor{lgrey}{HTML}{5ab4ac}
\definecolor{dgreen}{HTML}{005a32}
\definecolor{purple}{HTML}{ae017e}

\definecolor{editCol}{HTML}{0000FF}
\definecolor{maskCol}{HTML}{c51b7d}
\definecolor{lrColor}{HTML}{8856a7}
\definecolor{trColor}{HTML}{d01c8b}
\definecolor{ctColor}{HTML}{4dac26}
\definecolor{brickred}{HTML}{f03b20}

\definecolor{improveCol}{HTML}{4dac26}
\definecolor{worsenCol}{HTML}{d01c8b}

\definecolor{DarkBlue}{HTML}{00008B}
\definecolor{mscolor}{HTML}{01665e}
\definecolor{nmscolor}{HTML}{bf812d}
\definecolor{lgreen}{HTML}{ccece6}
\definecolor{dolive}{HTML}{308014}

\definecolor{lred}{HTML}{fbb4ae}
\definecolor{lblue}{HTML}{b3cde3}
\definecolor{lgreen}{HTML}{ccebc5}
\definecolor{lviolet}{HTML}{decbe4}
\definecolor{lorange}{HTML}{fed9a6}

\definecolor{lyellow}{HTML}{ffffcc}
\definecolor{lightgreen}{HTML}{97f8cd}

\colorlet{tablerowcolor4}{gray!50} 

\newcommand*{\textlabel}[2]{%
  \edef\@currentlabel{#1}
  \phantomsection
  #1\label{#2}
}

\colorlet{tableheadcolor}{gray!25} 
\colorlet{tablerowcolor}{gray!10} 
\colorlet{tablerowcolor2}{gray!45} 
\colorlet{tablerowcolor3}{gray!12} 

\newcommand{\rowcollight}{\rowcolor{tablerowcolor3}} %

\newcolumntype{a}{>{\columncolor{tablerowcolor}}r}
\definecolor{aicolor}{HTML}{018571}
\definecolor{occolor}{HTML}{ff7799}

\definecolor{aicolor}{HTML}{fc8d62}
\definecolor{occolor}{HTML}{253494}

\colorlet{tableheadcolor}{gray!25} 

\definecolor{neutralCol}{HTML}{dd1c77}
\definecolor{neutralGreen}{HTML}{31a354}
\definecolor{NewBlue}{HTML}{1879ba}
\definecolor{bleudefrance}{rgb}{0.19, 0.55, 0.91}  
\definecolor{AfTrColor}{HTML}{0868ac}  
\definecolor{BfTrColor}{HTML}{a8ddb5}  

\definecolor{AfCtColor}{HTML}{b10026}  
\definecolor{BfCtColor}{HTML}{fd8d3c}

\graphicspath{ {figures/} }

\newif{\ifhidecomments}
  \hidecommentsfalse 
\ifhidecomments
    \newcommand{\yueshen}[1]{}
    \newcommand{\vedant}[1]{}
    \newcommand{\koustuv}[1]{}
\else
    \newcommand{\yueshen}[1]{\textbf{\small\sffamily{\textcolor{green}{[#1 -- Yueshen]}}}}
    \newcommand{\vedant}[1]{\textbf{\small\sffamily{\textcolor{violet}{[#1 -- Vedant]}}}}
    \newcommand{\koustuv}[1]{\textbf{\small\sffamily{\textcolor{purple}{[#1 -- Koustuv]}}}}
  \fi
\newcommand{\edit}[1]{{#1}}

\newcommand{\para}[1]{\vspace{0.4em}\noindent\textbf{\textit{#1}~}}

\usepackage{booktabs}
\usepackage{tabularx}
\usepackage{lipsum} 
\usepackage{multirow}
\usepackage{makecell}
\usepackage{tikz}
\usepackage{pgfplots}
\pgfplotsset{compat=1.18}
\usepackage{pifont}

\usepackage{longtable}
\usepackage{placeins}

\usepackage[most]{tcolorbox}
\usepackage{listings}

\newtcblisting{promptbox}{
  listing only,
  breakable,
  enhanced,
  colback=gray!8,
  colframe=gray!40,
  boxrule=0.3pt,
  arc=2pt,
  left=6pt,
  right=6pt,
  top=6pt,
  bottom=6pt,
  width=\linewidth,
  before skip=6pt,
  after skip=6pt,
  listing options={
    basicstyle=\small\ttfamily,
    columns=fullflexible,
    keepspaces=true,
    breaklines=true,
    breakatwhitespace=false,
    breakautoindent=false,
    breakindent=0pt,
    postbreak=
  }
}

\title{TUX: Measuring Human--AI Tacit Understanding}

\author{
 \textbf{Yueshen Li\textsuperscript{1}},
 \textbf{Hanyi Min\textsuperscript{1}},
 \textbf{Vedant Das Swain\textsuperscript{2}},
 \textbf{Koustuv Saha\textsuperscript{1}}
\\
 \textsuperscript{1}University of Illinois Urbana-Champaign, \textsuperscript{2}New York University\\
 \{yueshen7, hanyimin, ksaha2\}@illinois.edu, v.das.swain@nyu.edu
}

\begin{document}
\maketitle

\begin{abstract}
As large language models (LLMs) increasingly act as collaborative partners, human--AI alignment is often evaluated through explicit task success, accuracy, or reward optimization. 
Yet many collaborative settings depend on tacit understanding: whether an agent can align with a human’s evaluative stance or representational priors without clear objectives, communication, or feedback. 
To study this capacity, we develop a spectrum-placement task inspired by the social party game Wavelength, in which humans and agents independently place concepts along subjective spectra. 
We operationalize the Tacit Understanding Index (TUX) as a pairwise \edit{behavioral measure} of similarity between human and agent judgments, and evaluate it with 241 human participants and 200 profile-conditioned LLM agents across four models. 
We find that nearest human--agent pairs in trait space achieve significantly higher TUX, suggesting that \edit{tacit alignment is associated with person-level characteristics rather than reflecting only random similarity}. 
Regression analyses show that TUX becomes more explainable as predictor sets become richer, with individual traits, decision-making styles, and confidence improving over aggregate trait-distance baselines. 
These findings suggest that \edit{TUX provides a measurable behavioral signal of human--LLM tacit understanding}, while revealing the limits of profile-based conditioning for capturing deeper representational alignment.
\end{abstract}

\section{Introduction}

As AI systems shift from tools to collaborative partners, a central challenge is not only whether AI can complete tasks, but whether humans and AI systems can become meaningfully aligned in shared situations.
Yet alignment in human--AI systems is commonly operationalized through explicit, goal-directed collaboration: agents are considered aligned when they jointly optimize a shared objective, complete a task efficiently, or maximize an explicit reward function~\cite{schmutz2024ai, fragiadakis2024evaluating}. 
That is, alignment is inferred through observable outcomes such as task accuracy, negotiation efficiency, or cooperative payoff~\cite{sidji2024human, schmutz2024ai, senoner2024explainable}.
While these measures are valuable, they capture only a limited form of collaboration.

Importantly, many forms of collaboration depend on something less visible: whether \edit{collaborators} develop a tacit sense of each other's perspective. 
Theories of \textit{common ground and grounding} argue that communication depends on the accumulation of shared assumptions about what participants know, believe, and take for granted~\cite{clark1991grounding}. 
In the case of human--human collaborations, people routinely rely on mutual understanding, shared expectations, and implicit calibration to coordinate action and make sense of each other's judgments.
Research on \textit{shared reality} suggests that people often seek not only agreement over external facts, but also a sense that another person shares their internal orientation toward the world---their interpretations, feelings, and evaluative stance~\cite{baek2022shared}. 
Relatedly, focal-point reasoning shows that people may coordinate around mutually salient interpretations even without direct communication~\cite{schelling1960strategy}. 
Together, these theories suggest that collaboration is not reducible to explicit instruction-following or task completion; it also involves representational alignment between minds. 

However, human--AI alignment has not yet fully incorporated the above tacit dimension. 
In fact, less is known about whether AI systems can approximate the implicit representational priors that make another agent's judgments feel understandable or socially attuned. 
This gap matters because many collaborative settings are not organized around a specific objective. 
In creative brainstorming, aesthetic judgment, interpersonal advising, and emotional companionship, a collaborator may need to infer what someone means by phrases such as ``this feels too cold,'' ``this seems trustworthy,'' or ``this does not fit my style.'' 
In these settings, success depends not only on producing a useful response, but also on whether the collaborator can recognize and align with the latent frames of reference through which the other person interprets the situation. 

This gap in human--AI collaborations has practical and theoretical consequences. 
Practically, an AI may appear successful because they produce fluent or high-performing outputs, while still failing to understand the user's implicit framing, values, or expectations. 
Theoretically, the absence of a tacit-understanding construct limits our ability to compare human--human and human--AI collaborations. 
Without a way to measure tacit alignment, we cannot distinguish between agents that merely optimize visible outcomes and agents that exhibit deeper representational calibration with human collaborators. 
Moreover, little empirical work examines how individual human traits shape alignment capacity, or whether LLMs exhibit stable behavioral signatures that influence tacit calibration~\cite{deng2024cooperativeness, park2024generative}.

In this paper, we ask whether tacit understanding between humans and AI can be measured as a form of representational calibration. Building on theories of common ground, shared reality, and individual-difference-based coordination~\cite{clark1991grounding,baek2022shared,schelling1960strategy,john1999big,frederick2005cognitive}, we test the hypothesis (H): \textbf{human--agent pairs with more similar trait-level descriptors would exhibit higher tacit understanding.}

To test this hypothesis, we use a controlled spectrum-placement task inspired by the social party game \textit{Wavelength} as a measurement setting for eliciting tacit representational judgments, where participants---both humans and profile-conditioned LLM agents---independently place cues along conceptual spectra.
The task removes explicit shared goals, direct communication, and performance feedback, allowing us to estimate several pair-level tacit understanding through participant's placements.
Accordingly, our work is guided by two research questions (RQs):

\para{RQ1:} How can tacit understanding be measured as a pair-level representational alignment construct in a weakly objective human--AI setting?

\para{RQ2:} How are personality and decision-making traits related to human--AI tacit understanding?

To address the above RQs, we conduct a user study with $N$=241 human participants and a bank of 200 profile-conditioned LLM agents instantiated across four model families (\texttt{Qwen3-14B}, \texttt{Qwen2.5-7B-Instruct}, \texttt{Mistral-Nemo}, and \texttt{Llama-3.1-8B}). 
Humans and agents complete the same spectrum-placement task under similar instruction, allowing us to compare independently generated judgements under weakly objective conditions. We compute pair-level tacit understanding using the Tacit Understanding Index (TUX), which measures similarity between human and agent placements across shared headers. 
We conduct regression modeling to examine how richer person-level descriptors---including personality traits, decision-making styles, and confidence---explain variation in TUX.

We find that nearest human--agent matches in trait space achieve significantly higher TUX than random matches (Cohen's $d_z$=0.27), supporting TUX as a measurable operationalization of tacit understanding.
Further, TUX is increasingly explainable as predictor sets become richer: models incorporating individual traits, decision-making styles, and confidence outperform a single aggregate trait-distance baseline. 
Finally, coefficient analyses suggest that \edit{TUX is associated more strongly with confidence in decisions and a subset of social and cognitive predictors}. 
These findings position tacit understanding as a measurable but partial form of human--AI representational alignment.

Our work makes the following contributions: 

\noindent \textbf{Operationalizing tacit understanding} through TUX, a pair-level behavioral measure of representational calibration between independent human and agent judgments.
    
\noindent \textbf{Trait--Behavior modeling of tacit understanding} by modeling pre-task trait with spectrum-placement behavior to examine how individual differences \edit{explain variation in TUX}.

\noindent\textbf{Human--Agent data Bank for tacit understanding}, for which we release the profile-conditioned LLM agent bank, persona profiles, agent responses, and task materials\footnote{\url{https://github.com/listenha/tux-data-bank}}, with de-identified human data to be released separately subject to the applicable data-sharing process, to support future research. 
\section{Related Work}\label{section:rw}

\para{Tacit Understanding and Shared Representation in Collaboration.}
Research in social science and coordination theory has long emphasized that successful collaboration often depends on tacit understanding~\cite{baek2022shared}. 
In classic accounts of tacit coordination, participants converge on mutually expected behaviors without negotiation by relying on salient focal points~\cite{schelling1960strategy}. 
A central mechanism underlying tacit coordination is common ground: the shared knowledge, beliefs, and assumptions that interlocutors treat as mutually recognized~\cite{clark1991grounding}. 
Common ground enables collaborators to interpret utterances, predict actions, and update expectations efficiently. 
These mechanisms explain how coordination can emerge even when communication is minimal
Further, theories of shared reality and intersubjectivity argue that individuals are motivated to align internal representations with others for mutual understanding~\cite{higgins2021shared, ehkirch2024understanding}. 
Such alignment may be grounded in shared cultural background, common experiences, social values, or worldviews~\cite{kahan2011cultural, wang2021roles}. 
Recent work further highlights the value of community-centered data for developing models that better reflect the values and expectations of the communities they are intended to support~\cite{goyal2026vastu,kim2026llumi}.%
Prior work characterized tacit knowledge in graphic design as experience-based, context-sensitive, and often embedded in situated design actions~\cite{son2024demystifying}. 
Building on the above body of research, our work refers to \emph{tacit understanding} as the ability of humans and agents to align with one another's latent representational priors without communication, shared reward, or feedback. 

\para{Persona Conditioning in LLMs.}
As LLMs are increasingly deployed as social and collaborative agents, substantial effort has been devoted to conditioning models on persona or profile information to induce behavioral characteristics~\cite{chen2024persona,shimgekar2026ai, kim2026pair}. 
A common approach is direct persona prompting, in which the model is instructed to ``act as'' a particular character or adopt specific traits~\cite{tseng2024two, chen2024persona}. 
However, prior work suggests that such shifts may not reliably translate into consistent behavioral changes across tasks~\cite{mannekote2025belief, zou2024selfreport}. 
Other approaches attempt to induce persona shifts by manipulating internal model representations, such as through vector steering or activation editing~\cite{turner2023steering, qiu2024spectral}, or by fine-tuning models on persona-specific data~\cite{wozniak2024personalized}. 
Recent generative-agent work offers an alternative: structured contextual grounding through persistent memory and profile construction. 
These systems elicit richer characteristics through extended interaction and maintain them through memory architectures that retrieve task-relevant contextual fragments during inference~\cite{park2023generative, park2024generative}. 
Recent work has proposed a knowledge ecosystem in which users create, share, and inject knowledge modules into LLM interactions~\cite{zhao2025knoll}, and how a chatbot mediates organizational memory by retrieving knowledge embedded in communication practices~\cite{lee2026choir}. 
Further, personality and cognitive traits influence how people perceive social situations, anticipate others, and coordinate behavior~\cite{john1999big, frederick2005cognitive, deng2024cooperativeness, wang2021roles, duddu2026not, das2024teacher}. 
However, in the case of human--AI collaborations, most methods optimize toward population-level preference consistency, as in reinforcement learning from human feedback, direct preference optimization, and related preference-based training paradigms~\cite{christiano2017deep, rafailov2023direct, liu2025dpo}. 
This paper combines these two perspectives. We construct persona-conditioned LLM agents with reusable profile information, and we pair this with human trait profiling to examine whether person-level descriptors explain tacit understanding. 
We evaluate whether profile-conditioned agents can participate in a non-goal-oriented calibration task where the outcome is representational alignment.

\begin{figure*}[t]
    \centering
    \includegraphics[width=0.95\textwidth]{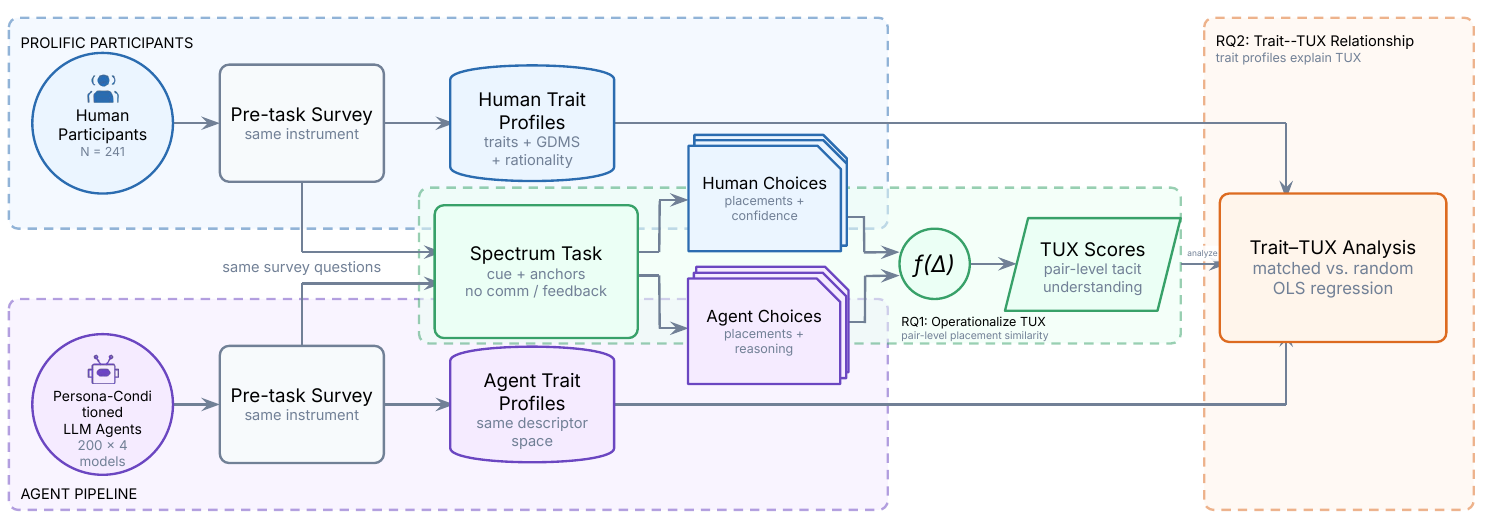}
    \caption{A schematic figure of our study design: Human participants and profile-conditioned LLM agents complete the same pre-task survey instrument and the same spectrum-placement task. Survey responses define human and agent trait profiles, while task responses define human and agent choices.} 
    \label{fig:study_design_workflow}
\end{figure*}

\para{Human--AI Collaboration Benchmarks.}
Research has introduced a variety of benchmarks for studying human--AI cooperation. 
CoBlock~\cite{wu2024coblock} is a multi-agent 3D blocks-world platform where human and LLM agents manipulate objects and communicate to achieve cooperative goals. 
The Constrained Human--AI Cooperation challenge~\cite{du2024chaic} evaluates embodied social intelligence, requiring an AI helper to infer a physically constrained human's intent and limitations while completing everyday tasks. 
Cooperative game settings have also been adapted:
Codenames has been used as a testbed where an LLM codemaster gives clues to guide a human guesser under partial information~\cite{sidji2024human}, multi-agent suites such as SPIN-Bench~\cite{yao2025spin} and Concordia~\cite{smith2025concordia} evaluate planning, negotiation, social intelligence, trust, and mixed-motive interaction.
Relatedly, human-AI interactions and alignment are explained by the mutual theory of mind~\cite{wang2021towards,pal2026we} and mental models~\cite{bansal2019beyond}.
Beyond benchmark settings, recent work has also examined LLM-based collaborators for workplace settings~\cite{das2025ai,mysore2025prototypical}.
While these benchmarks are valuable for testing social reasoning under interactive and strategically complex conditions, they typically retain explicit objectives, communication channels, shared rewards, or feedback loops that scaffold coordination. 
As a result, successful performance may reflect task-specific heuristics, following conversational signals, or optimizing a visible payoff rather than aligning with a partner's latent internal framing. 
Our work complements existing benchmarks by minimizing explicit scaffolding and foregrounding partner diversity. 
Our setting asks whether humans and agents can independently produce similar representations based on their underlying traits.
This shifts the evaluation target from overt task success to tacit representational fit, providing a controlled setting for measuring when and how human--AI tacit understanding emerges.

\section{Study Design and Methods}\label{sec:data}

\autoref{fig:study_design_workflow} summarizes our study design. 
Human participants and profile-conditioned LLM agents complete the same pre-task survey instrument and spectrum-placement task. Survey responses define human and agent trait profiles, while task responses define human and agent choices. 
Pairwise placement similarity measures TUX, which is then analyzed against trait profiles through matched-versus-random comparison and regression.

\subsection{Spectrum-Placement Task}

\para{Wavelength Game.} Our task is inspired by the social party game \emph{Wavelength}, a cooperative guessing game in which players coordinate around meanings along a conceptual spectrum~\cite{morrison2025wavelength}.
In the original game, players are shown a conceptual spectrum defined by two opposing anchors, such as ``hot--cold,'' ``formal--casual,'' or ``safe--risky.''
One player serves as the cue-giver and privately sees a hidden target location somewhere along the spectrum. 
The cue-giver then provides a verbal clue intended to help the other players infer the target placement, without directly naming a number or placement.
For example, if the spectrum is ``hot--cold'' and the hidden target is closer to the ``hot'' side, the cue-giver might say ``fresh coffee.''
The remaining players act as guessers: they interpret the clue and place a pointer on the spectrum where they believe the hidden target is located. 
Success is scored by how close their placement is to the hidden target. 
The game, therefore, rewards not only descriptive precision, but also shared conceptual framing: players must have similar assumptions about how examples map into a spectrum. 

\begin{figure}[t]
    \centering
    \includegraphics[width=0.8\columnwidth]{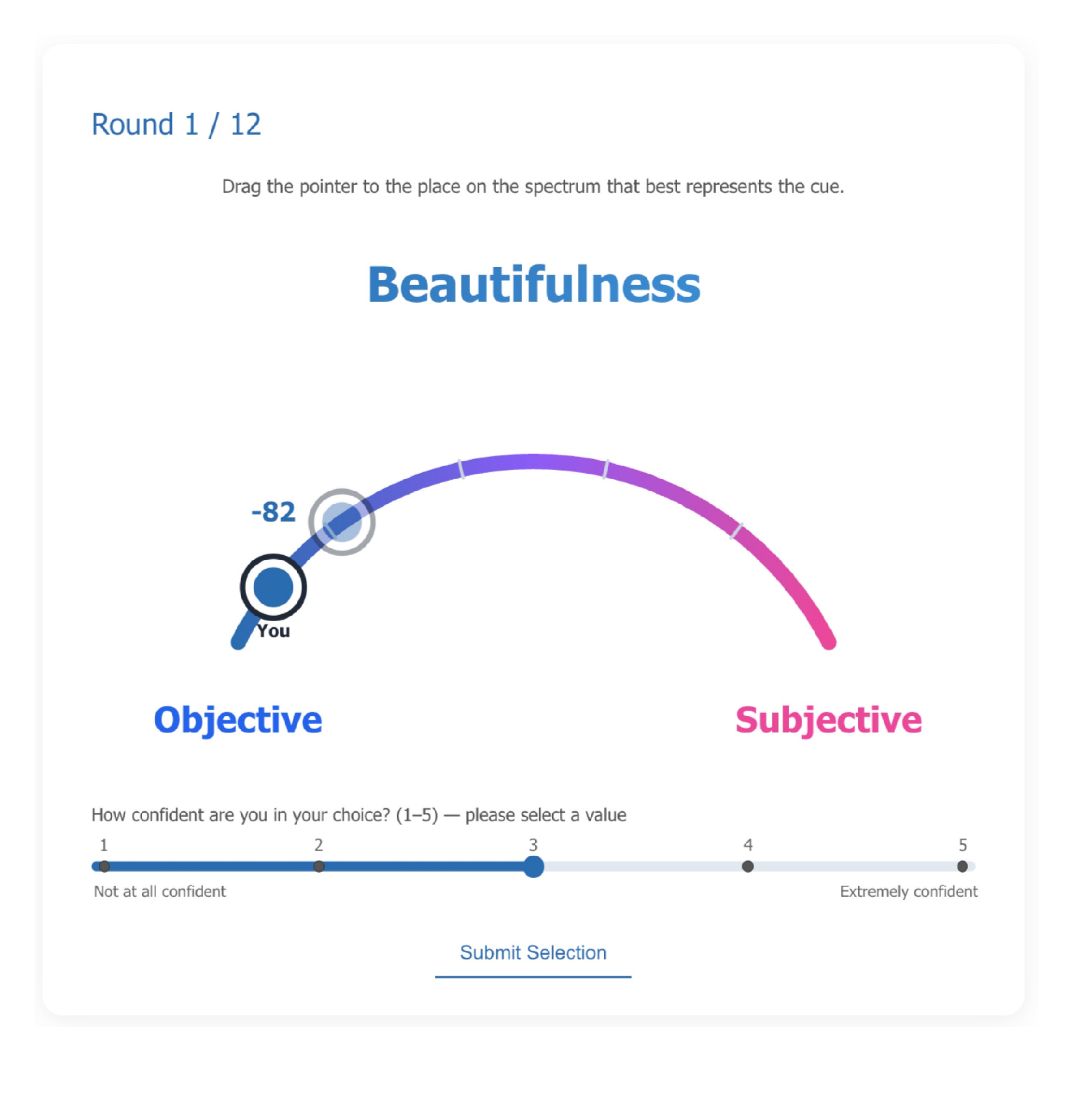}
    \caption{Interface of the spectrum placement task.}
    \label{fig:game_interface}
\end{figure}

In our study, to operationalize tacit alignment without explicit communication or feedback, we adapt this structure into a symmetric single-response task, implemented as the interface shown in~\autoref{fig:game_interface}. 
Each participant completes an independent session as a ``guesser.'' 
Participants receive spectrum-cue headers that are predefined by the researchers. 
They are told that the target location has been selected by a like-minded peer, and their task is to estimate where they believe that peer would place the pointer along the spectrum.
\edit{We note that this ``like-minded peer'' framing was used for human participants only to encourage more careful reasoning; LLM agents were instead instructed to place each cue based on their own personality, experiences, and perspective.}

This design removes overt coordination and direct feedback. 
Participants respond independently and never see partner responses during the task. 
The perceived task frame remains collaborative---``align with a similar mind''---but without outcome feedback, making the task a controlled probe of tacit representational calibration.

\para{Spectrum Headers.} Our task consists of a number of \emph{spectrum headers}. Each header defines a conceptual continuum through two anchors (e.g., \textit{Subjective}--\textit{Objective}) and a cue (e.g., \textit{Beautifulness}). 
We treat these headers as measurement items: each header must be coherent enough that participants understand the scale, but open-ended for capturing differences in priors and values.

Our goal in constructing the headers was to elicit structured variation in interpretation. 
We manually curated anchor pairs and cue items according to three criteria: 1) \textbf{Coherent anchor contrast:} the two endpoints define an interpretable conceptual continuum; 2) \textbf{Interpretive openness:} the cue admits multiple plausible placements rather than a single obvious answer; and 3) \textbf{Domain breadth:} the full header set spans semantic, affective, social, and judgment-oriented domains.

Under this design, individuals with similar priors or perspectives would likely tend to produce closer placements, while those with divergent priors may locate the same cue differently. 
This makes header responses useful behavioral traces for estimating tacit understanding through pair-level calibration.~\autoref{tab:example_headers} shows the fixed headers.

\begin{table}[t]
\centering
\footnotesize
\setlength{\tabcolsep}{3pt}
\resizebox{\columnwidth}{!}{\begin{tabular}{p{0.4\columnwidth}p{0.28\columnwidth}p{0.28\columnwidth}}
\textbf{Cue} & \textbf{Left Anchor} & \textbf{Right Anchor} \\
\toprule
A coworker takes credit for your idea & Confront them directly & Let it go to avoid tension \\
\rowcollight A large party & Leave early and rest & Stay late and meet people \\
Planning a weekend trip & Detailed itinerary & Go with the flow \\
\rowcollight Quitting stable job for a new startup & Uncomfortable & Excited \\
Bungee jumping & Stressful & Thrilling \\
\rowcollight Beach & Calm & Crowded \\
\end{tabular}}
\caption{Example spectrum headers used in our task. \edit{These six headers formed the fixed core set used for the primary analysis.} Each header combines a cue with a pair of conceptual anchors, designed to preserve semantic coherence while allowing interpretive variation.}
\label{tab:example_headers}
\end{table}

\subsection{Evaluating Tacit Alignment}

To evaluate tacit alignment across human and language model (LM) agents, we constructed a dual-track experiment that \edit{mirrors response modality and task structure across human and AI agent participants}. 
Both groups first complete individual-difference questionnaires, followed by the same spectrum alignment tasks under matched instruction. The goal is to assess behavioral calibration---how well a participant's placements align with that of a similarly profiled peer---under minimal communication. 
We investigate how individual traits are associated with independent judgment alignments between humans and agents.
A key design decision is the inclusion of a large, structured agent bank with reusable memory profiles, enabling controlled human--agent comparison and trait--behavior analysis.

\subsubsection{Human Participants} 
We recruited human participants through the Prolific platform. 
Participants were compensated at an average rate of \$10.04 per hour, and the median completion time was 11 minutes and 57 seconds. 
Each participant's session consisted of two phases: a pre-task trait survey and the spectrum alignment task.
All participants provided informed consent and completed the study via a hosted Qualtrics interface. 
Participants were included only if they passed all three sanity-check questions---two in the pre-task survey and one in the spectrum alignment game---and spent at least three minutes on the game task. 
After filtering out participants who failed attention check questions, the final retained dataset contains 241 human participants. 

\edit{The retained participants had a mean age of 37.72 years ($SD=11.24$, range 19--65). Prolific-recorded sex was 52.7\% female (127/241), 46.9\% male (113/241), and 0.4\% prefer not to say (1/241). Simplified ethnicity was 63.9\% White, 13.3\% Black, 10.4\% Asian, 7.1\% Mixed, 4.6\% Other, and 0.8\% prefer not to say. All retained participants reported U.S. nationality, and 94.6\% resided in the United States.}

\para{Pre-task Survey.} 
Participants completed a survey questionnaire designed to elicit psychological and decision-making traits. This included: 1) Big Five Personality Inventory (BFI-10)~\cite{rammstedt2007measuring}---measuring openness, conscientiousness, extraversion, agreeableness, and neuroticism, 2) Empathy Scale--Cognitive Subscale (BES-A)~\cite{carre2013basic}---assessing cognitive perspective-taking and awareness of others' emotional states, 3) Rational Ability Subscale of the Rational--Experiential Inventory (REI)~\cite{pacini1999relation}---measuring logical analysis and reasoning skill, 4) Narcissism Subscale of the Short Dark Triad (SD3)~\cite{jones2014introducing}---evaluating self-perceived social dominance and attention-seeking tendencies, and 5) General Decision-Making Style (GDMS)~\cite{scott1995decision}---capturing rational, intuitive, avoidant, dependent, and spontaneous decision tendencies.

\para{Spectrum Alignment Task.}
Participants then played the spectrum game described above. Responses were recorded on a continuous slider (-100 to +100), with randomized non-core headers and a shared set of core headers.

\para{Data Collection and Scale.} The task used 22 spectrum headers in total. Human participants answered a subset of headers, at an average of 11.71 headers per participant: the shared core headers were administered broadly across participants, while additional non-core headers were sampled from the remaining header pool. Participants also reported confidence on a 1--5 scale for each round.

\subsubsection{Agent Bank Construction}
To enable controlled human--agent comparison, we constructed a bank of 200 profile-conditioned LLM agents. Each agent began with a demographic seed profile adapted from an open-source generative society simulation repository~\cite{park2024generative}. Because these seed profiles provide only limited context for inducing distinctive and diverse agent behavior, we further used an extended, simulated \emph{profile-enrichment semi-structured interview script} to elicit richer autobiographical content from each seed persona. 

\edit{The agent bank was constructed independently of the 241 human participants. During profile enrichment, interview questions were answered sequentially: each autobiographical response was stored in the agent's memory stream and became available as context for subsequent interview questions, progressively enriching the profile from its demographic seed.}

\begin{figure*}[t]
    \centering
    \includegraphics[width=\textwidth]{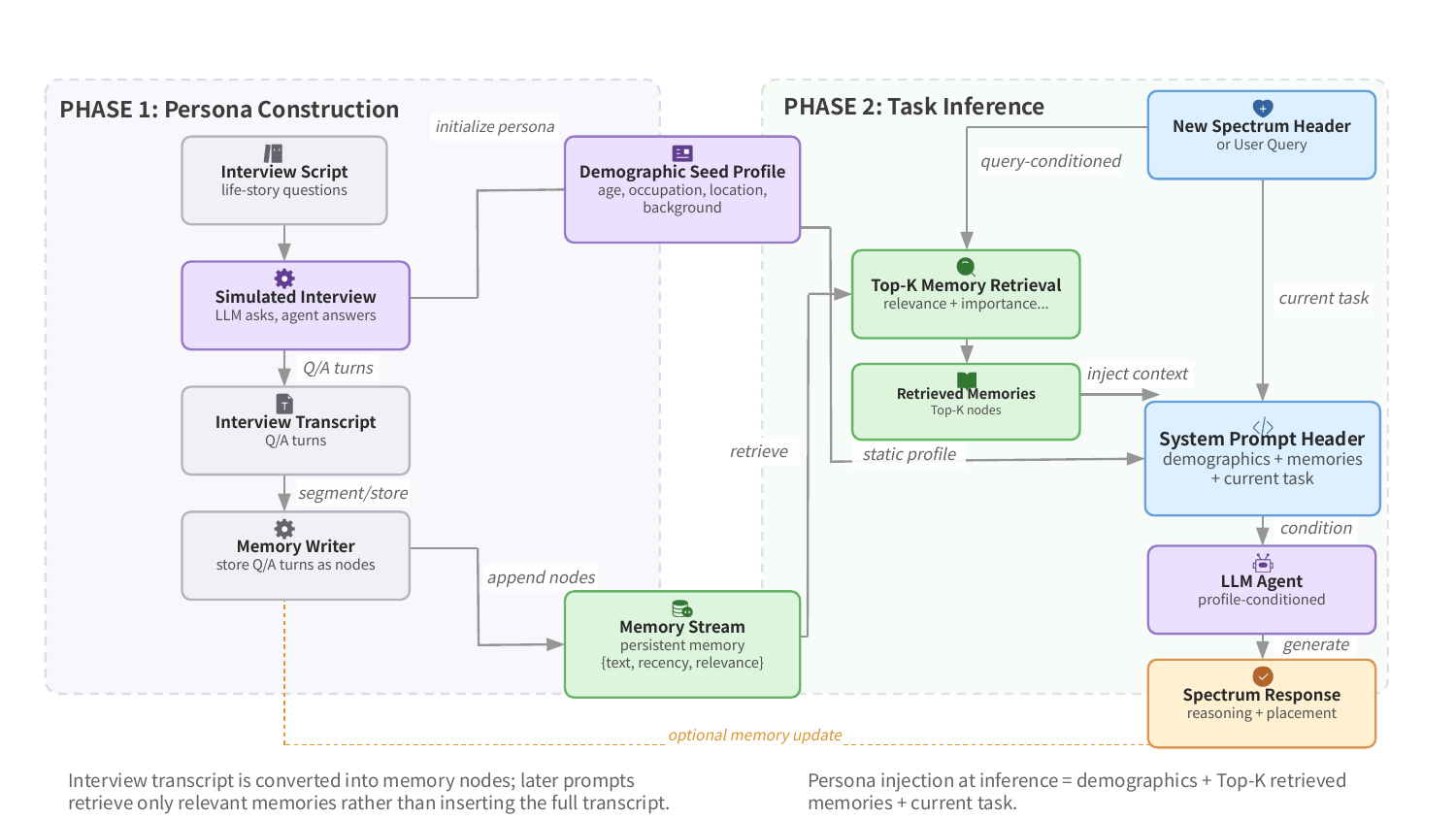}
    \caption{Generative-agent persona and memory workflow. Demographic seed profiles initialize a simulated interview that produces autobiographical Q/A transcripts. These transcripts are segmented into persistent memory nodes rather than inserted wholesale into later prompts. During spectrum-placement inference, the current header retrieves top-$K$ relevant memories, which are combined with demographic context into a system prompt header that conditions the LLM agent's response.}
    \label{fig:agent_memory_arch}
\end{figure*}

Agents were not given the full profile-enrichment transcript, and each task query triggered a top-$K$ retrieval over the memory stream, selecting relevant memories based on semantic similarity, importance, and recency. 
Retrieved memory nodes were combined with demographic context and the current task header to construct the system prompt. 
This approach preserves rich persona context while avoiding context overflow. \edit{\autoref{fig:agent_memory_arch} summarizes the persona-construction and memory-retrieval workflow; additional construction materials are provided in Appendix~\ref{app:agent_persona_materials}.}

\para{Behavioral inference across models.}
We instantiated the same 200 agent profiles across four models: \textit{Qwen3-14B}, \textit{Qwen2.5-7B-Instruct}, \textit{Mistral-Nemo}, and \textit{Llama-3.1-8B}. Each model received the same demographic seed profile, retrieved memory context, spectrum headers, and task instructions, and responded in zero-shot mode without fine-tuning or reward optimization. Unless otherwise noted, the main analyses use \texttt{Qwen3-14B} as the primary family because it was complete in the canonical data, showed the highest repeated-attempt consistency, had the strongest average trait--header expression summary, and produced the largest matched-versus-random TUX gap. The remaining families are reported as robustness analyses in Appendix~\ref{app:agent_family_ablation}. 
\section{Results}

\subsection{RQ1: Measuring TUX}

A core goal of this work is to operationalize tacit understanding in a weakly objective setting using a measurable pair-level quantity. For each human--agent pair and each overlapping observed header $j$, we define per-header similarity as $\mathtt{s_{haj} = 1 - \lvert y_{hj} - y_{aj} \rvert}$, where $\mathtt{y_{hj}, y_{aj} \in [0,1]}$ are the normalized human and agent choices on header $j$.

\edit{We use the fixed six-header core subset as our primary TUX definition because it was designed for uniform coverage across participants. Of the 241 retained participants, 233 completed all six core headers; the small amount of missing core data comes from several initial Prolific trial runs used to verify the study workflow end-to-end. Let $C_h$ denote the observed core headers for human $h$. We define
$\mathrm{TUX}^{\mathrm{core}}_{ha} = \frac{1}{|C_h|}\sum_{j \in C_h} s_{haj}$.
Larger TUX indicates greater human--agent similarity.}

Unless otherwise noted, our analyses use \texttt{Qwen3-14B} as the primary agent family, which showed the highest repeated-attempt consistency among the four agent families ($\mathtt{ICC}$=0.72). Appendix~\ref{app:agent_family_ablation} reports robustness analyses across all four models and shows that the trait--TUX regression pattern is not unique to this primary family.

To test whether this operationalization captures meaningful tacit alignment, we compare the TUX of the nearest human--agent match in the 8-trait space against a random baseline defined as the mean TUX over all candidate agent attempts.
\edit{The matched mean TUX is 0.6670, compared with 0.6507 for the random baseline, with a statistically significant paired $t$-test ($t(240)=4.24$, $p<0.001$) and a modest effect size (Cohen's $d_z=0.27$).}
Within this primary family, these results indicate that the proposed TUX operationalization captures a statistically reliable, though modest, aspect of tacit human--agent similarity.

\edit{As a complementary broader-coverage measure, we also compute $\mathrm{TUX}^{\mathrm{all}}_{ha}$ by averaging over all headers observed for each human. This produces the same qualitative matched-versus-random pattern (0.6699 vs.\ 0.6577; $d_z=0.29$), while allowing header composition to vary across participants.}

\begin{figure}[t]
\centering
\includegraphics[width=0.85\linewidth]{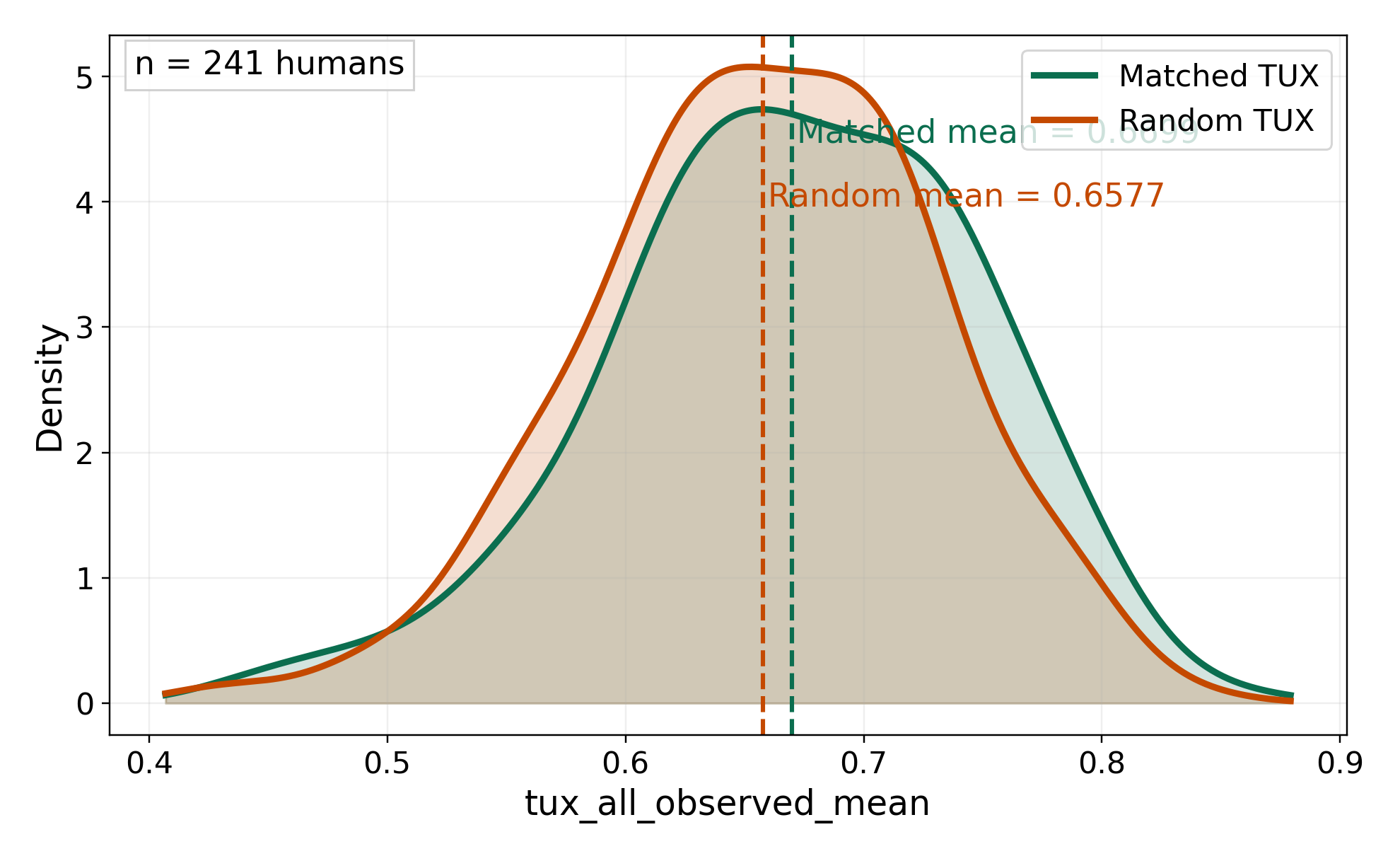}
\caption{\edit{Matched-versus-random distributions for the primary core-header TUX analysis.}}
\label{fig:matched_random_tux_density}
\end{figure}

\subsection{RQ2: Explaining TUX with Traits}

We fit OLS regression models predicting $\mathrm{TUX}^{\mathrm{core}}_{ha}$ from increasingly rich human--agent descriptor gaps.
In the broad Model~4, the predictor set contains eight primary trait gaps---extraversion, agreeableness, conscientiousness, neuroticism, openness, cognitive empathy, rational ability, and narcissism---five pre-task GDMS gaps---rational, intuitive, avoidant, dependent, and spontaneous---and the human's mean confidence on the shared core headers. The model can be written compactly as

\begin{equation}
\vspace{-0.5em}
\mathrm{TUX}^{\mathrm{core}}_{ha}
=
\alpha
+
\sum_{k=1}^{8}\beta_k d^{(k)}_{ha}
+
\sum_{\ell=1}^{5}\gamma_\ell g^{(\ell)}_{ha}
+
\delta c^{\mathrm{core}}_{h}
+
\varepsilon_{ha}.
\vspace{-0.5em}
\end{equation}

Here, $d^{(k)}_{ha}$ denotes one of the eight primary trait gaps, $g^{(\ell)}_{ha}$ denotes one of the five GDMS gaps, each gap is an absolute human--agent difference in pooled-$z$ standardized space, and $c^{\mathrm{core}}_{h}$ is the human's mean confidence on the shared core headers. Model~1 uses a single aggregated 8-trait distance. Model~2 replaces this aggregate with the eight individual primary trait gaps. Model~3 adds the five pre-task GDMS gaps. Model~4 adds mean confidence on the shared core headers.

As shown in~\autoref{tab:tux_full_fit}, explanatory power increases monotonically as the predictor set becomes richer.
\edit{Full-dataset $R^2$ rises from 0.07 in Model~1 to 0.19 in Model~4, and the corresponding Pearson correlation between fitted and observed TUX rises from $r$=0.26 to $r$=0.44.}
This progression suggests that TUX is not merely a noisy proxy: it becomes increasingly explainable as more concentrated person-level information is included. In other words, the latent ingredients of tacit understanding are not exhausted by one aggregate trait distance, and richer predictor sets recover more of the variation in the pair-level TUX signal.

The same pattern appears in predictive sensitivity, also reported in~\autoref{tab:tux_full_fit}.
\edit{Under repeated $k$-fold ($k$=5) human-level cross-validation splits, mean SMAPE changes from 12.01\% in Model~1 to 11.43\% in Model~4, while mean training $R^2$ improves from 0.07 to 0.20. When held-out predictions are pooled across all five splits, Pearson correlation improves from $r$=0.27 in Model~1 to $r$=0.40 in Model~4.}
Thus, the richer models do not merely fit the full dataset more closely; they also improve held-out predictive performance under repeated human-level splits.

\begin{table}[t]
\centering
\scriptsize
\setlength{\tabcolsep}{2pt}
\resizebox{\columnwidth}{!}{
\begin{tabular}{lr@{}lr@{}lr@{}lr@{}l}
\textbf{Predictor} & \multicolumn{2}{c}{\textbf{M1}} & \multicolumn{2}{c}{\textbf{M2}} & \multicolumn{2}{c}{\textbf{M3}} & \multicolumn{2}{c}{\textbf{M4}} \\
\toprule
\rowcollight \multicolumn{9}{l}{\emph{Aggregated distance}} \\
Agg. 8-trait distance & -0.2626 & *** & -- &  & -- & & -- & \\
\hdashline
\rowcollight \multicolumn{9}{l}{\emph{Primary trait gaps}} \\
Extraversion gap & -- && -0.2178&*** & -0.2154&*** & -0.1791&*** \\
Agreeableness gap & -- && -0.0107&*** & -0.0152&*** & -0.0405&*** \\
Conscientiousness gap & -- && 0.0062&* & 0.0063&* & 0.0131&*** \\
Neuroticism gap & -- && -0.0159&*** & -0.0125&*** & -0.0200&*** \\
Openness gap & -- && -0.0046& & -0.0020& & 0.0017& \\
Cognitive empathy gap & -- && -0.1276&*** & -0.1067&*** & -0.0616&*** \\
Rational ability gap & -- && -0.0342&*** & -0.0262&*** & 0.0014& \\
Narcissism gap & -- && -0.1229&*** & -0.1234&*** & -0.1177&*** \\
\hdashline
\rowcollight \multicolumn{9}{l}{\emph{GDMS gaps}} \\
GDMS rational gap & -- && -- && -0.0517&*** & -0.0240&*** \\
GDMS intuitive gap & -- && -- && -0.0236&*** & 0.0009& \\
GDMS avoidant gap & -- && -- && 0.0739&*** & 0.0350&*** \\
GDMS dependent gap & -- && -- && -0.0131&*** & -0.0052&* \\
GDMS spontaneous gap & -- && -- && 0.0740&*** & 0.0722&*** \\
\hdashline
\multicolumn{9}{l}{\emph{Confidence}} \\
Mean confidence, core headers & -- && -- && -- && -0.2877&*** \\
\hdashline
\rowcollight \multicolumn{9}{l}{\emph{Model-level metrics}} \\
Full-data $R^2$ & 0.0690&& 0.1058&& 0.1206&& 0.1943&\\
Held-out SMAPE & 12.0100&& 11.7818&& 11.8278&& 11.4339&\\
Held-out Pearson's $r$ & 0.2680&*** & 0.3194&*** & 0.3169&*** & 0.4030&***\\
\bottomrule
\end{tabular}}
\caption{\edit{OLS regression models with standardized coefficients.} SMAPE and Pearson's $r$ are computed by pooling held-out predictions across $k$-fold splits ($k$=5) (* $p$<0.05, ** $p$<0.01, *** $p$<0.001).}
\label{tab:tux_full_fit}
\end{table}

\edit{\para{Broader-coverage comparison.}
Using the all-observed mean TUX as the target yields the same overall progression, but with somewhat lower fit: full-dataset $R^2$ rises from 0.05 in Model~1 to 0.15 in Model~4, compared with 0.07 to 0.19 for the core target; fitted--observed Pearson correlation increases from $r$=0.23 to $r$=0.39, compared with $r$=0.26 to $r$=0.44 for the core target.}

\para{Comparing Predictors.}
Beyond overall model fit, the coefficient structure shows that some human--agent differences matter much more than others.
To make predictors directly comparable, we inspect standardized coefficients from the broad model that includes the eight trait distances, five GDMS distances, and mean confidence on the shared core headers.
\autoref{tab:tux_full_fit} reports the standardized coefficients across the model sequence.

\edit{The dominant predictor is human mean core confidence, which enters with a negative coefficient ($\beta$=-0.29). Among the primary trait gaps, the largest coefficients in magnitude in Model~4 are extraversion ($\beta$=-0.18), narcissism ($\beta$=-0.12), and cognitive empathy ($\beta$=-0.06).}
This pattern suggests that tacit understanding is not uniformly distributed across the full trait set, but instead is driven disproportionately by a smaller subset of social and cognitive predictors.

The negative confidence effect is particularly notable. Follow-up diagnostics suggest that confidence is partly linked to response extremity: more confident participants tended to make more extreme or idiosyncratic placements, which were less likely to align with agent placements and were therefore associated with lower TUX.

\para{Qualitative interpretation of the dominant predictors.}
\edit{Beyond confidence, the strongest primary trait predictors are extraversion, narcissism, and cognitive empathy, which together point to a plausible decomposition of tacit understanding into \textit{social expressivity}, \textit{self oriented interpersonal stance}, and \textit{deliberative collaboration}. A cautious interpretation is that extraversion may matter because it shapes how legible or consistently projected a person's choices are to a partner; narcissism may matter because stronger self-oriented or competitive interpersonal tendencies can reduce supplementary fit and make tacit matching harder.}

These interpretations are also broadly consistent with prior work linking personality and cognitive traits to social perception, cooperation, and team performance~\cite{peeters2006team,kichuk1997big,lynch2022way}. 
\section{Discussion  and Conclusion}\label{section:discussion}

We designed a spectrum-placement task for studying tacit understanding between humans and AI under weakly objective conditions, and proposed TUX as a pair-level measure. 
Using independent placements by humans ($N$=241) and profile-conditioned agents ($N$=200), we found that nearest human--agent matches in trait space achieve significantly higher TUX than a random baseline, supporting TUX as a credible measure of one aspect of tacit understanding. 
Further, richer descriptor sets improved the explanation and prediction of TUX, while a smaller subset of predictors carries much of the signal. Together, these results suggest that tacit understanding is measurable and structurally constrained, but not fully captured by simple compatibility heuristics. More broadly, they show that human--AI alignment should not be studied only through explicit goal achievement: representational fit itself can be quantified, compared, and modeled.

\para{Operationalizing Tacit Understandings.}
Our results suggest that tacit understanding in human--AI settings can be operationalized as a measurable behavioral construct rather than treated as a purely qualitative intuition. 
The key contribution of this work is not only the spectrum placement task, but also an operationalizable pair-level TUX that makes tacit calibration observable and statistically testable. 
In weakly objective settings, where explicit task success is unavailable, this kind of operationalization is important: it provides a way to study whether two agents independently converge on similar internal framings.
This has implications for alignment evaluation: human--AI collaboration should not be assessed only through external performance, but also through whether systems can approximate the latent evaluative frames through which humans interpret ambiguous situations.

\para{Tacit Understanding is Structured but Partial.}
Our findings suggest that tacit understanding is real but partial. Nearest human--agent matches in the active 8-trait space achieved significantly higher TUX than a random baseline, indicating that human--agent tacit understanding is not reducible to chance. 
However, the absolute gap remains modest, and even the richest descriptor models explain only a limited share of TUX variance. 
We view this combination of reliability and incompleteness as theoretically meaningful. 
The implication is that profile-conditioned agents may approximate some aspects of human representational alignment, but current profiles remain too shallow to fully capture the tacit priors that shape human judgment.

\para{Designing Richer Measures of Person-Level Alignment.}
The regression progression provides a complementary implication. As predictor sets become richer, both explanatory fit and predictive sensitivity improve. This pattern supports the interpretation that TUX is a meaningful pair-level construct: it responds systematically to psychologically and behaviorally relevant descriptors.
For benchmark design, this suggests that future evaluations of tacit understanding can invest in richer person descriptors, including traits, states, preferences, lived experiences, and contextual styles, beyond what we considered in this study. 
The same lesson appears from the header-specific analysis: broader header sets carry substantially richer trait signal than the shared core subset alone, implying that future benchmarks may benefit from both richer \emph{person descriptions} and richer \emph{prompt sets}.

\section{Limitations and Future Directions}

This work has limitations, which also suggest interesting future directions. 
First, TUX is a starting point \edit{for} operationalizing tacit understanding. 
While the matched-versus-random result supports it as a meaningful behavioral measure, tacit understanding is a broader construct than pairwise placement similarity in a single task family. 
Our task is built around a spectrum-placement paradigm inspired by Wavelength. 
This structure is useful for isolating internal calibration without explicit communication, but it also constrains the form that tacit understanding can take. 
Future work can examine whether the same conclusions hold under alternative operationalizations, including dialogic, affective, and co-creative settings.
\edit{Because the human task used a ``like-minded peer'' framing whereas agents responded from their own profile-conditioned perspective, future work can also test a human control condition without the peer framing.}

Also, the current descriptor set is not comprehensive. 
Our work likely explains only a limited share of TUX variance, and inspires future directions of richer descriptors and contextual factors. 
These may include lived experience, cultural background, aesthetic preference, domain familiarity, and other forms of contextual prior not captured by standard survey scales. 
Future benchmarks can expand both the breadth and the granularity of person descriptors.

Further, the task header coverage is unequal. 
\edit{Our primary analysis uses the fixed core-header subset to improve comparability across participants, while the broader 22-header analyses include unequal header coverage across participants.}
\edit{Future studies with denser, fully shared header coverage would allow cleaner comparisons while retaining broader prompt diversity.}

Further, the agent bank depends on a specific persona-construction pipeline. 
Our agents are not generic language models; they are profile-conditioned agents built from demographic seeds, interview-style enrichment, and memory retrieval. 
This is a strength for controlled experimentation, but it also means the findings should be interpreted as applying to this class of profile-conditioned LLM agents rather than to all AI systems. 
Different profile construction or conditioning methods may yield different tacit-understanding patterns.
Additionally, our model-family comparison remains bounded. Although the study includes multiple models and a shared agent bank, it is still only a limited slice of the broader space of contemporary language models. Future work can test whether the present structural patterns replicate across other larger families, instruction-tuned variants, and non-LLM agent architectures.

Finally, our analyses are explanatory and predictive, but not causal. 
The fact that richer descriptor sets improve TUX prediction does not imply that any individual trait directly causes tacit understanding. 
Likewise, the coefficient ranking should be interpreted as a structured association pattern rather than a definitive mechanism. 
Our work inspires future research in exploring causal pathways through targeted manipulations or intervention-based experimental designs. 
\section{Ethical Considerations}
\label{sec:ethics}

This study was reviewed and approved by the authors' institutional review board (IRB) at the University of Illinois Urbana-Champaign. 
Human participants were recruited through Prolific and completed the study through a hosted Qualtrics interface after providing informed consent. Participants were compensated at an average rate of \$10.04 per hour, and participation was voluntary. We applied pre-specified quality-control criteria, including sanity-check questions and minimum task-duration filtering, to ensure data validity without penalizing attentive participants.

The human dataset was de-identified before analysis. We do not report participant names, contact information, or other directly identifying information. Survey responses were used only to construct aggregate trait descriptors for analysis, and individual-level results are reported only in aggregate. Because the task asks participants to make subjective judgments and report confidence, we treat these responses as behavioral measurements rather than evaluations of participant ability or correctness.

The agent portion of the study uses profile-conditioned LLM agents constructed from demographic seed profiles and simulated interview memories. These profiles are used to support controlled experimental comparison, not to make claims about real individuals or demographic groups. We therefore interpret agent behavior as the behavior of model-conditioned personas rather than as representative of any real population. Any public release of the data bank will exclude personally identifying human information and will document the synthetic nature and limitations of the agent profiles. 
\section{AI Involvement Disclosure}
\label{sec:ai_disclosure}

AI-assisted tools were used only for language editing, formatting assistance, and improving clarity during manuscript preparation. The study design, data collection, statistical analyses, interpretation of results, and experimental decisions were conducted by the authors.

\section{Acknowledgments}
This work was partly supported by the Center for Social \& Behavioral Studies (CSBS) Small Grant Program at the University of Illinois Urbana-Champaign.

\bibliography{0paperACL}

\appendix
\clearpage
\appendix

\makeatletter
\setlength{\@fptop}{0pt}
\setlength{\@fpsep}{10pt}
\setlength{\@fpbot}{0pt plus 1fil}

\setlength{\@dblfptop}{0pt}
\setlength{\@dblfpsep}{10pt}
\setlength{\@dblfpbot}{0pt plus 1fil}
\makeatother

\setcounter{table}{0}
\setcounter{figure}{0}
\counterwithin{table}{section}
\counterwithin{figure}{section}
\renewcommand{\thetable}{\thesection.\arabic{table}}
\renewcommand{\thefigure}{\thesection.\arabic{figure}}

\section{Study Materials}
\label{app:study_materials}

This appendix summarizes the study materials used to collect human and agent responses. We include the pre-task survey instruments, the full set of spectrum headers, \edit{the human and agent task instructions,} and the profile/interview structure used to instantiate persona-conditioned agents.

\subsection{Pre-task Survey Instruments}
\label{app:pretask_survey}

Before the spectrum-placement task, participants completed a pre-task survey measuring personality, empathy, rational ability, narcissism, and decision-making style. Unless otherwise noted, all items used a five-point Likert scale from 1 = \emph{Disagree strongly} to 5 = \emph{Agree strongly}. Table~\ref{tab:pretask_instruments} summarizes the constructs used in the survey.

\begin{table*}[t]
\centering
\small
\caption{Pre-task survey instruments used to construct human and agent descriptors.}
\label{tab:pretask_instruments}
\resizebox{\textwidth}{!}{
\begin{tabular}{p{0.18\linewidth}p{0.18\linewidth}p{0.48\linewidth}p{0.08\linewidth}}
Construct & Source & Items used & Scale \\
\toprule
Big Five personality & BFI-10 \citep{rammstedt2007measuring} &
Reserved; generally trusting; tends to be lazy; relaxed, handles stress well; has few artistic interests; outgoing, sociable; tends to find fault with others; does a thorough job; gets nervous easily; has an active imagination. & 1--5 \\
\hline
Cognitive empathy & BES-A cognitive empathy subscale \citep{carre2013basic} &
Finds it hard to know when friends are frightened; can usually work out when friends are scared; can often understand how people are feeling before they say; not usually aware of friends' feelings; has trouble figuring out when friends are happy. & 1--5 \\
\hline
Rational ability & REI rational ability subscale \citep{pacini1999relation} &
Not good at complicated problems; not good at problems requiring careful logical analysis; not very analytical; reasoning things out carefully is not a strong point; does not reason well under pressure. & 1--5 \\
\hline
Narcissism & SD3 narcissism subscale \citep{jones2014introducing} &
People see me as a natural leader; I hate being the center of attention (reverse-coded); many group activities are dull without me; I know I am special because everyone keeps telling me so; I like to get acquainted with important people. & 1--5 \\
\hline
Decision-making style & GDMS \citep{scott1995decision} &
Logical and systematic decision-making; avoiding decisions until pressure is on; consulting other people before important decisions; relying on intuition; making snap decisions; exploring all options before deciding. & 1--5 \\
\hline
\end{tabular}}
\end{table*}

\subsection{Full Spectrum-Placement Header Set}
\label{app:full_headers}

Each spectrum-placement trial presents a cue and two spectrum anchors. Human participants place the cue along the spectrum according to where they believe a like-minded peer would place it. \edit{LLM agents instead place the same cue according to their own profile-conditioned personality, experiences, and perspective.} Table~\ref{tab:full_headers} lists the full set of 22 headers used in the study.

\begin{table*}[t]
\centering
\small
\caption{Full set of spectrum-placement headers. \edit{IDs 0--5 form the fixed core set used for the primary analysis.}}
\label{tab:full_headers}
\begin{tabular}{rlll}
\hline
ID & Cue & Left anchor & Right anchor \\
\hline
0 & A coworker takes credit for your idea & Confront them directly & Let it go to avoid tension \\
1 & A large party & Leave early and rest & Stay late and meet people \\
2 & Planning a weekend trip & Detailed itinerary & Go with the flow \\
3 & Quitting stable job for a new startup & Uncomfortable & Excited \\
4 & Bungee jumping & Stressful & Thrilling \\
5 & Beach & Calm & Crowded \\
6 & Your music taste & Quiet & Bombastic \\
7 & Public speaking & Uncomfortable & Energizing \\
8 & Smartphone & Useful & Harmful \\
9 & Alarm Clock & Helpful & Annoying \\
10 & Neighbor & Friendly & Nosy \\
11 & Beautifulness & Objective & Subjective \\
12 & Vacation & Familiar place & New destination \\
13 & Drinking & Cringe & Cool \\
14 & Competition & Motivating & Stressful \\
15 & Glass & Durable & Fragile \\
16 & Aircraft engine & Distraction & Focus \\
17 & Coffee & Relaxing & Energizing \\
18 & Surprise party & Overwhelming & Delightful \\
19 & Texting style & Brief and direct & Expressive and detailed \\
20 & Genius & Lucky & Hardworking \\
21 & Deadline & Early completion & Last-minute burst \\
\hline
\end{tabular}
\end{table*}

\subsection{Human and Agent Task Instructions}
\label{app:task_instructions}

\edit{The human and agent conditions used the same cue--anchor spectrum structure but different task framing. In particular, the ``like-minded peer'' instruction was used only for human participants. Agents were instead instructed to place each cue according to their own profile-conditioned personality, experiences, and perspective. We reproduce the task instructions below to make this distinction explicit.}

\paragraph{Human instructions.}
Before beginning the spectrum-placement task, human participants received the following instruction:

\begin{quote}
\textbf{Important instructions}

Please carefully consider where you place your marker.

In this study, your responses will later be compared with another participant whose background and survey responses are highly similar to yours---someone who thinks much like you do. For now, you can think of this person as another version of yourself. Your goal is to place the marker as your ``other self'' would---at the precise location you believe they would choose. Small percentage differences matter.

\textbf{Quality checks are applied.} Extremely fast or non-deliberate responses may not be eligible for payment.
\end{quote}

\paragraph{Agent spectrum-task prompt.}
For each agent trial, the model received its profile-conditioned context followed by the task template below:

\begin{quote}
\small
\textbf{Task:} You are playing a Wavelength-style game.

A spectrum is defined between two opposites:

\texttt{[LEFT] --- [RIGHT]}

A clue is given: \texttt{[CUE]}.

Based on your personality, experiences, and perspective, place a marker on the spectrum (0--100) where you think the clue belongs. 0 means it's completely on the \texttt{[LEFT]} side, 100 means it's completely on the \texttt{[RIGHT]} side.

As you answer, I want you to take the following steps:

\textbf{Step 1)} Describe what each end of the spectrum represents for this clue. What would it mean for the clue to be at 0 (completely \texttt{[LEFT]})? What would it mean for it to be at 100 (completely \texttt{[RIGHT]})? (``Range Interpretation'')

\textbf{Step 2)} Write a few sentences explaining your reasoning for where you place the marker on this spectrum. Consider your personal experiences, values, and perspective. (``Reasoning'')

\textbf{Step 3)} Provide your numerical response (0--100) based on your reasoning. Use your intuitive judgment---don't overthink it. (``Response'')

Here is the game header:

\textbf{Spectrum:} \texttt{[LEFT] --- [RIGHT]}

\textbf{Clue:} \texttt{[CUE]}

Where on this spectrum (0--100) does the clue belong?

\textbf{Range:} [0, 100]
\end{quote}

\edit{The agent returned a structured response containing its interpretation of the spectrum endpoints, reasoning, and a single numerical placement. The task instruction itself did not ask the agent to infer or imitate a ``like-minded peer.''} \edit{Although humans responded on a $[-100,100]$ slider and agents on a $[0,100]$ numerical scale, both were linearly normalized to $[0,1]$ before computing TUX.}

\subsection{Agent Persona Construction Materials}
\label{app:agent_persona_materials}

Agents were initialized from structured demographic seed profiles and then enriched through a simulated semi-structured interview. The seed profile provided stable background attributes, while the interview generated autobiographical memory content used for retrieval-conditioned downstream responses. \edit{The agent bank was constructed independently of the 241 human participants in this study. During enrichment, interview questions were answered sequentially: each autobiographical response was stored in the memory stream and became available as context for subsequent interview questions, progressively enriching the profile from its demographic seed.}

\paragraph{Seed profile schema.}
Each seed profile contained demographic and background fields that served as static identity anchors. The schema included name fields, age, sex, race/ethnicity descriptors, city/state, family background, education, work status, marital status, language background, financial background, and a short profile summary. For privacy and reproducibility, we report the schema categories rather than a complete raw seed profile. Table~\ref{tab:agent_profile_schema} summarizes the fields used.

\begin{table*}[t]
\centering
\small
\caption{Agent seed profile schema. We report field categories rather than raw profile values to avoid exposing sensitive or identifying example details.}
\label{tab:agent_profile_schema}
\begin{tabular}{p{0.22\linewidth}p{0.68\linewidth}}
\hline
Category & Fields \\
\hline
Basic identity & first name, last name, age, sex, city, state \\
Race and ethnicity & ethnicity, race, detailed race, Hispanic origin \\
Family background & residence at age 16, same residence since age 16, family structure at age 16, family income at age 16, parents' education, mother's work history \\
Education and work & highest degree received, work status, occupation-related summary when available \\
Household and social status & marital status, military service, citizenship status, language background \\
Financial background & total wealth / financial-status field \\
Profile summary & short natural-language summary of the agent's background, values, and life context \\
\hline
\end{tabular}
\end{table*}

\paragraph{Interview script.}
The simulated interview contained 45 open-ended questions designed to elicit autobiographical memories, values, daily routines, relationships, health, stress, finances, work, family background, and future goals. Each question--answer turn was stored as a memory node and used as part of the agent's persistent memory stream. During task inference, relevant memory nodes were retrieved and included in the agent prompt.

Table~\ref{tab:interview_domains} summarizes the major interview domains, and Table~\ref{tab:interview_examples} gives representative questions. The full 45-question script is included in the supplemental repository.

\begin{table}[t]
\centering
\small
\caption{Major domains covered by the agent interview script.}
\label{tab:interview_domains}
\begin{tabular}{ll}
\hline
Domain & Question IDs \\
\hline
Life history and turning points & 1--4 \\
Family and close relationships & 5--7 \\
Neighborhood and living environment & 8--10 \\
Daily routines and responsibilities & 11--15 \\
Health and wellbeing & 16--22 \\
Religion, social media, and coping & 23--26 \\
Finances and household budget & 27--32 \\
Work and workplace context & 33--40 \\
Parents' work background & 41--43 \\
Future goals and values & 44--45 \\
\hline
\end{tabular}
\end{table}

\begin{table*}[t]
\centering
\small
\caption{Representative questions from the 45-question simulated interview script.}
\label{tab:interview_examples}
\begin{tabular}{rp{0.82\linewidth}}
\hline
ID & Question \\
\hline
1 & Tell me the story of your life, from childhood to education, family, relationships, and major life events. \\
2 & Was there a crossroads in your life where multiple paths were available and your choice made a significant difference in defining who you are? \\
5 & Tell me more about family who are important to you. Do you have a partner or children? \\
8 & Tell me about the neighborhood and area in which you are living now. \\
11 & Across a typical week, how do your days vary? \\
16 & Tell me all about your health. \\
20 & Tell me about how you have been doing over the past year. \\
27 & Tell me about how you make ends meet and what the monthly budget looks like for you and your family. \\
33 & What is or was your occupation, and what kind of work do you do? \\
40 & Overall, how do you feel about your financial situation? \\
44 & Imagine yourself a few years from now. What do you hope your life will look like? \\
45 & What do you value the most in your life? \\
\hline
\end{tabular}
\end{table*}

\subsection{Agent Memory Architecture}
\label{app:agent_architecture}

Figure~\ref{fig:agent_memory_arch} illustrates the persona-injection and memory-stream workflow used for profile-conditioned agents. Each agent begins from a demographic seed profile. A simulated interview then elicits life-history and value-related responses, which are segmented and written into a persistent memory stream. During spectrum-placement inference, the current header retrieves top-$K$ relevant memory nodes, which are combined with demographic context and the current task into a system prompt header. The LLM agent then generates a spectrum-placement response consisting of reasoning and a numerical placement.

\section{Additional TUX Robustness and Diagnostic Analyses}
\label{app:tux_diagnostics}

\subsection{Separate Human and Agent Descriptor Regression}
\label{app:separate_traits_regression}

\edit{As a complementary analysis using the broader all-observed TUX target,} we model TUX using human--agent descriptor gaps and additionally fit models using human and agent descriptors as separate predictors. This specification asks whether TUX is predictable from the directional positions of the human and agent in descriptor space, rather than from absolute gaps alone. A-family models use separate human and agent descriptors; B-family models additionally include same-trait human$\times$agent interactions.

\begin{table*}[t]
\centering
\scriptsize
\caption{Full-dataset OLS regression extension models predicting \edit{all-observed} TUX using human and agent descriptors as separate predictors. Predictor cells report signed standardized coefficients with significance stars. A-family models use separate human and agent descriptors; B-family models additionally include same-trait human$\times$agent interactions. Significance legend: * $p<0.05$, ** $p<0.01$, *** $p<0.001$.}
\label{tab:separate_traits_extension_full_regression}
\begin{tabular}{lcccccc}
\hline
Predictor / metric & A1 & A2 & A3 & B1 & B2 & B3 \\
\hline
\multicolumn{7}{l}{\emph{Human primary traits}} \\
Human Extraversion & $0.3011^{***}$ & $0.2892^{***}$ & $0.2449^{***}$ & $0.2820^{***}$ & $0.2702^{***}$ & $0.2259^{***}$ \\
Human Agreeableness & $-0.0100^{***}$ & $-0.0117^{***}$ & 0.0006 & $-0.0222^{***}$ & $-0.0239^{***}$ & $-0.0116^{***}$ \\
Human Conscientiousness & 0.0031 & $0.0383^{***}$ & $0.0363^{***}$ & 0.0017 & $0.0369^{***}$ & $0.0349^{***}$ \\
Human Neuroticism & $0.0516^{***}$ & $0.0136^{***}$ & $-0.0207^{***}$ & $0.0528^{***}$ & $0.0147^{***}$ & $-0.0196^{***}$ \\
Human Openness & 0.0031 & $-0.0201^{***}$ & $-0.0107^{***}$ & 0.0034 & $-0.0198^{***}$ & $-0.0104^{***}$ \\
Human Cognitive empathy & $-0.1641^{***}$ & $-0.1460^{***}$ & $-0.0919^{***}$ & $-0.1641^{***}$ & $-0.1462^{***}$ & $-0.0921^{***}$ \\
Human Rational ability & $0.0240^{***}$ & $0.0448^{***}$ & $0.0353^{***}$ & $0.0169^{***}$ & $0.0378^{***}$ & $0.0283^{***}$ \\
Human Narcissism & $0.0124^{***}$ & $0.0335^{***}$ & $0.0215^{***}$ & $0.0085^{**}$ & $0.0296^{***}$ & $0.0176^{***}$ \\
\hline
\multicolumn{7}{l}{\emph{Agent primary traits}} \\
Agent Extraversion & $-0.0339^{***}$ & $-0.0266^{***}$ & $-0.0266^{***}$ & $-0.0124^{***}$ & -0.0051 & -0.0051 \\
Agent Agreeableness & $-0.0129^{***}$ & -0.0052 & -0.0052 & -0.0037 & 0.0040 & 0.0040 \\
Agent Conscientiousness & $0.0158^{***}$ & $0.0132^{***}$ & $0.0132^{***}$ & $0.0197^{***}$ & $0.0172^{***}$ & $0.0172^{***}$ \\
Agent Neuroticism & -0.0015 & $-0.0055^{*}$ & $-0.0055^{*}$ & -0.0030 & $-0.0071^{*}$ & $-0.0071^{**}$ \\
Agent Openness & -0.0031 & -0.0003 & -0.0003 & -0.0035 & -0.0007 & -0.0007 \\
Agent Cognitive empathy & -0.0033 & -0.0044 & -0.0044 & -0.0032 & -0.0041 & -0.0041 \\
Agent Rational ability & $0.0364^{***}$ & $0.0306^{***}$ & $0.0306^{***}$ & $0.0501^{***}$ & $0.0441^{***}$ & $0.0441^{***}$ \\
Agent Narcissism & $-0.0137^{***}$ & $-0.0095^{***}$ & $-0.0095^{***}$ & $-0.0095^{***}$ & -0.0054 & $-0.0054^{*}$ \\
\hline
\multicolumn{7}{l}{\emph{Human GDMS}} \\
Human GDMS rational & -- & $-0.1389^{***}$ & $-0.1029^{***}$ & -- & $-0.1384^{***}$ & $-0.1023^{***}$ \\
Human GDMS avoidant & -- & $0.0810^{***}$ & $0.0537^{***}$ & -- & $0.0807^{***}$ & $0.0535^{***}$ \\
Human GDMS dependent & -- & $0.0100^{***}$ & $0.0162^{***}$ & -- & $0.0108^{***}$ & $0.0171^{***}$ \\
Human GDMS intuitive & -- & $-0.0180^{***}$ & $-0.0112^{***}$ & -- & $-0.0174^{***}$ & $-0.0106^{***}$ \\
Human GDMS spontaneous & -- & $-0.0589^{***}$ & $-0.0291^{***}$ & -- & $-0.0551^{***}$ & $-0.0253^{***}$ \\
\hline
\multicolumn{7}{l}{\emph{Agent GDMS}} \\
Agent GDMS rational & -- & $-0.0216^{***}$ & $-0.0216^{***}$ & -- & $-0.0230^{***}$ & $-0.0230^{***}$ \\
Agent GDMS avoidant & -- & 0.0012 & 0.0012 & -- & 0.0015 & 0.0015 \\
Agent GDMS dependent & -- & $-0.0096^{***}$ & $-0.0096^{***}$ & -- & $-0.0115^{***}$ & $-0.0115^{***}$ \\
Agent GDMS intuitive & -- & $-0.0109^{***}$ & $-0.0109^{***}$ & -- & $-0.0118^{***}$ & $-0.0118^{***}$ \\
Agent GDMS spontaneous & -- & $-0.0264^{***}$ & $-0.0264^{***}$ & -- & $-0.0321^{***}$ & $-0.0321^{***}$ \\
\hline
\multicolumn{7}{l}{\emph{Same-trait interactions}} \\
Extraversion H$\times$A & -- & -- & -- & $0.0504^{***}$ & $0.0502^{***}$ & $0.0502^{***}$ \\
Agreeableness H$\times$A & -- & -- & -- & $0.0201^{***}$ & $0.0201^{***}$ & $0.0201^{***}$ \\
Conscientiousness H$\times$A & -- & -- & -- & $-0.0124^{***}$ & $-0.0127^{***}$ & $-0.0127^{***}$ \\
Neuroticism H$\times$A & -- & -- & -- & $0.0132^{***}$ & $0.0131^{***}$ & $0.0131^{***}$ \\
Openness H$\times$A & -- & -- & -- & 0.0026 & 0.0027 & 0.0027 \\
Cognitive empathy H$\times$A & -- & -- & -- & -0.0002 & -0.0005 & -0.0005 \\
Rational ability H$\times$A & -- & -- & -- & $-0.0215^{***}$ & $-0.0211^{***}$ & $-0.0211^{***}$ \\
Narcissism H$\times$A & -- & -- & -- & $0.0140^{***}$ & $0.0139^{***}$ & $0.0139^{***}$ \\
\hline
\multicolumn{7}{l}{\emph{Same-GDMS interactions}} \\
GDMS rational H$\times$A & -- & -- & -- & -- & 0.0036 & 0.0036 \\
GDMS avoidant H$\times$A & -- & -- & -- & -- & -0.0009 & -0.0009 \\
GDMS dependent H$\times$A & -- & -- & -- & -- & -0.0037 & -0.0037 \\
GDMS intuitive H$\times$A & -- & -- & -- & -- & 0.0031 & 0.0031 \\
GDMS spontaneous H$\times$A & -- & -- & -- & -- & $0.0122^{***}$ & $0.0122^{***}$ \\
\hline
\multicolumn{7}{l}{\emph{Confidence}} \\
Mean confidence, core headers & -- & -- & $-0.2356^{***}$ & -- & -- & $-0.2356^{***}$ \\
\hline
\multicolumn{7}{l}{\emph{Model-level metrics}} \\
Full-dataset $R^2$ & 0.1120 & 0.1341 & 0.1804 & 0.1150 & 0.1373 & 0.1835 \\
Full-data Pearson & $0.3346^{***}$ & $0.3662^{***}$ & $0.4247^{***}$ & $0.3391^{***}$ & $0.3705^{***}$ & $0.4284^{***}$ \\
Full-data Spearman & $0.3279^{***}$ & $0.3583^{***}$ & $0.4087^{***}$ & $0.3322^{***}$ & $0.3626^{***}$ & $0.4128^{***}$ \\
\hline
\end{tabular}
\end{table*}

\begin{table*}[t]
\centering
\small
\caption{Predictive sensitivity for separate human/agent trait regression extension models under repeated 80/20 human-level train/test splits using the \edit{all-observed TUX target}. SMAPE is averaged across five held-out test folds; Pearson and Spearman are computed by pooling held-out predictions across all five splits. $R^2_{\mathrm{train}}$ is averaged across the corresponding training folds. Significance legend: * $p<0.05$, ** $p<0.01$, *** $p<0.001$.}
\label{tab:separate_traits_extension_predictive}
\begin{tabular}{llcccc}
\hline
Model & Predictor set & SMAPE & Pearson & Spearman & $R^2_{\mathrm{train}}$ \\
\hline
A1 & Human + agent primary traits & 9.2880 & $0.3263^{***}$ & $0.3090^{***}$ & 0.1040 \\
A2 & A1 + human/agent GDMS & 9.2939 & $0.3246^{***}$ & $0.2983^{***}$ & 0.1266 \\
A3 & A2 + core confidence & 9.1692 & $0.3753^{***}$ & $0.3551^{***}$ & 0.1767 \\
\hline
B1 & A1 + same-trait interactions & 9.2681 & $0.3315^{***}$ & $0.3149^{***}$ & 0.1069 \\
B2 & B1 + GDMS values/interactions & 9.2746 & $0.3296^{***}$ & $0.3027^{***}$ & 0.1297 \\
B3 & B2 + core confidence & 9.1515 & $0.3796^{***}$ & $0.3586^{***}$ & 0.1798 \\
\hline
\end{tabular}
\end{table*}

The separate-descriptor extension provides two diagnostic observations. First, modeling human and agent descriptors separately improves fit relative to the \edit{all-observed} gap-based model: the best separate-descriptor model reaches $R^2=0.1835$ ($r=0.4284$, $\rho=0.4128$), compared with $R^2=0.1524$ ($r=0.3904$, $\rho=0.3614$) for the \edit{all-observed} gap-based model. Second, most of this gain comes from separate human/agent descriptors and core confidence rather than interaction terms: adding same-trait interactions increases full-dataset $R^2$ only from $0.1804$ to $0.1835$ when confidence is included. We therefore treat this as a diagnostic result: TUX is partly predictable from directional human and agent profiles, but the \edit{core-header gap-based model in the main text remains our primary analysis}.

\subsection{Agent Family Ablation}
\label{app:agent_family_ablation}

The main results use \texttt{Qwen3-14B} as the primary agent family. Table~\ref{tab:agent_family_reliability} summarizes reliability and response-distribution diagnostics. \texttt{Qwen3-14B} had the highest repeated-attempt ICC ($0.7235$), the highest binned response entropy ($2.7642$), and the strongest average trait--header expression summary (mean absolute trait--header correlation $=0.0909$). We therefore treat \texttt{Qwen3-14B} as a pragmatic primary family because it provides the strongest combination of repeated-attempt consistency, non-collapsed response diversity, trait-conditioned expressivity, and matched-versus-random TUX signal.

\begin{table*}[t]
\centering
\small
\caption{Agent-family reliability and response-distribution diagnostics. All families contain 200 profiles, 22 headers, and complete parsed placements.}
\label{tab:agent_family_reliability}
\begin{tabular}{lccccc}
\hline
Agent family & Parse rate & Within SD & ICC & Header SD & Entropy \\
\hline
\texttt{Qwen3-14B} & 1.0000 & 0.0769 & 0.7235 & 0.1190 & 2.7642 \\
\texttt{Qwen2.5-7B} & 1.0000 & 0.0688 & 0.3875 & 0.0825 & 1.9864 \\
\texttt{Mistral-Nemo} & 1.0000 & 0.1339 & 0.4786 & 0.1735 & 2.5470 \\
\texttt{Llama-3.1-8B} & 1.0000 & 0.1230 & 0.3452 & 0.1523 & 1.9847 \\
\hline
\end{tabular}
\end{table*}

Table~\ref{tab:agent_family_model4} compares the regression specification across families \edit{using the primary core-header TUX target}. The Model~4 specification includes eight primary trait gaps, five pre-task GDMS gaps, and human mean confidence. The trait--TUX relationship is not unique to \texttt{Qwen3-14B}: all four families yield significant fitted--observed correlations and significant pooled held-out correlations. Interestingly, \texttt{Qwen2.5-7B} achieves the strongest full-data and held-out regression metrics \edit{($R^2=0.2493$, held-out $r=0.4755^{***}$)}, indicating that the regression result is robust across agent families even though the primary family was selected for broader reliability and matched-alignment criteria.

\begin{table*}[t]
\centering
\small
\caption{Final Model~4 \edit{core-header TUX} regression comparison across agent families. Full-data Pearson and Spearman are fitted--observed correlations on the full dataset. Held-out Pearson and Spearman are computed by pooling held-out predictions across five human-level train/test splits. Significance legend: $^{***}p<0.001$.}
\label{tab:agent_family_model4}
\begin{tabular}{lccccccc}
\hline
Agent family & Full $R^2$ & Full $r$ & Full $\rho$ & SMAPE & Held-out $r$ & Held-out $\rho$ & Train $R^2$ \\
\hline
\texttt{Qwen3-14B}
& 0.1943
& $0.4407^{***}$
& $0.4257^{***}$
& 11.4339
& $0.4030^{***}$
& $0.3880^{***}$
& 0.1979 \\

\texttt{Qwen2.5-7B}
& 0.2493
& $0.4993^{***}$
& $0.5035^{***}$
& 12.0200
& $0.4755^{***}$
& $0.4972^{***}$
& 0.2465 \\

\texttt{Mistral-Nemo}
& 0.1956
& $0.4423^{***}$
& $0.4376^{***}$
& 12.9140
& $0.4412^{***}$
& $0.4386^{***}$
& 0.1885 \\

\texttt{Llama-3.1-8B}
& 0.1875
& $0.4330^{***}$
& $0.4273^{***}$
& 11.4640
& $0.4436^{***}$
& $0.4481^{***}$
& 0.1802 \\
\hline
\end{tabular}
\end{table*}

Table~\ref{tab:agent_family_matched_random} reports matched-versus-random \edit{core-header TUX} comparisons across families. \texttt{Qwen3-14B} produces the largest matched-minus-random gap ($0.0163$; $t=4.2435$, $d_z=0.2733$), followed by \texttt{Llama-3.1-8B} ($0.0121$; $t=2.8924$, $d_z=0.1863$). \edit{All four model families show statistically significant positive matched-minus-random gaps, although the effect sizes vary across families.} These results suggest that model-family choice affects the strength of the matched-versus-random TUX signal.

\begin{table*}[t]
\centering
\small
\caption{Matched-versus-random \edit{core-header TUX} comparison across agent families. Matched agents are nearest neighbors in the active 8-trait space; random TUX is the mean over candidate agent attempts for each human.}
\label{tab:agent_family_matched_random}
\begin{tabular}{lcccccc}
\hline
Agent family & Matched & Random & Gap & Rel. improv. & $t$ & $d_z$ \\
\hline
\texttt{Qwen3-14B}
& 0.6670
& 0.6507
& 0.0163
& 2.5103\%
& $4.2435^{***}$
& 0.2733 \\

\texttt{Qwen2.5-7B}
& 0.6400
& 0.6357
& 0.0044
& 0.6845\%
& $2.0725^{*}$
& 0.1335 \\

\texttt{Mistral-Nemo}
& 0.6504
& 0.6410
& 0.0094
& 1.4733\%
& $2.4757^{*}$
& 0.1595 \\

\texttt{Llama-3.1-8B}
& 0.6739
& 0.6618
& 0.0121
& 1.8236\%
& $2.8924^{**}$
& 0.1863 \\
\hline
\end{tabular}
\end{table*}

\section{When TUX Matters}
\label{app:when_tux_matters}

This section reports diagnostics that ask when the matched-versus-random TUX advantage is largest and which headers appear most informative for benchmark construction.

\subsection{Header-Level TUX Component Gaps}
\label{app:header_tux_gap}

To examine where the matched-versus-random TUX advantage is concentrated, we decomposed the human-level TUX gap into per-header components. For each header $j$, we computed the signed component gap
\begin{equation}
\begin{aligned}
g_j
&=
\frac{1}{|H_j|}\sum_{h \in H_j}
\Bigg[
1 - |y_{hj} - y_{m(h)j}| \\
&\qquad\qquad
-
\frac{1}{N_a}\sum_{a}
\left(1 - |y_{hj} - y_{aj}|\right)
\Bigg].
\end{aligned}
\end{equation}
where $H_j$ is the set of humans who answered header $j$, $m(h)$ is the trait-matched agent for human $h$, and $N_a$ is the number of candidate agent attempts. Positive values indicate that the trait-matched agent is closer to the human response than the random candidate-pool average on that header.

\begin{figure*}[!t]
\centering
\includegraphics[width=0.72\textwidth]{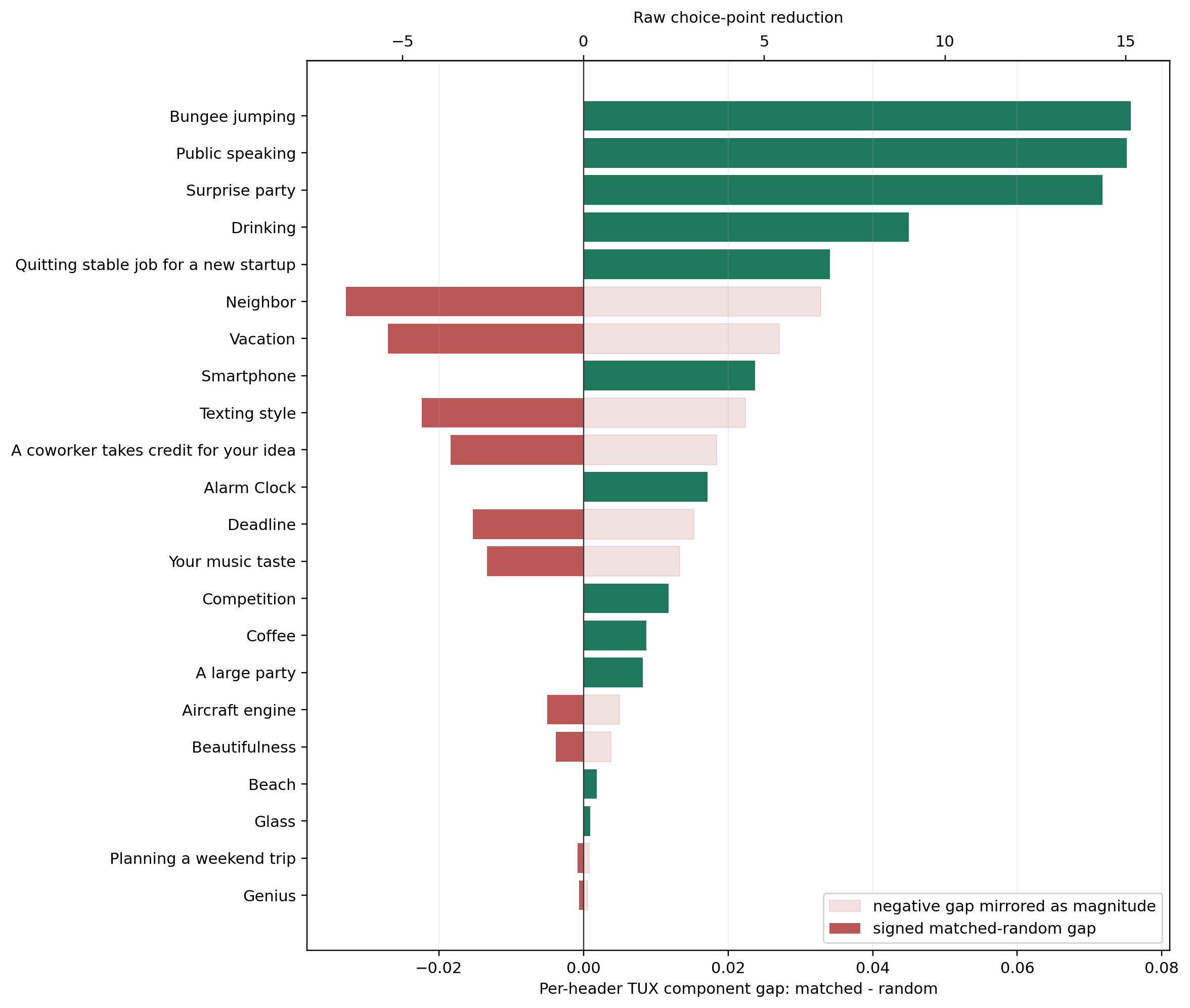}
\caption{Per-header matched-versus-random TUX component gaps. Positive values indicate headers where the trait-matched agent is closer to the human response than the random candidate-pool average. The top axis maps normalized TUX component gaps to raw slider-point reductions on the original $[-100,100]$ response scale.}
\label{fig:header_tux_component_gap}
\end{figure*}

The largest positive component gaps appeared on headers such as \textit{Bungee jumping} ($g_j=0.0757$), \textit{Public speaking} ($g_j=0.0751$), and \textit{Surprise party} ($g_j=0.0718$). Because normalized TUX distances are defined on a $[0,1]$ scale mapped from the original $[-100,100]$ slider, these correspond to average raw-spectrum reductions of approximately 15.14, 15.02, and 14.36 points, respectively (Figure~\ref{fig:header_tux_component_gap}).

We also examined whether header-level response variance predicted where the matched advantage was largest. Human response variance was not clearly associated with the absolute component gap ($r=0.2607$, $p=0.2412$; $\rho=0.1237$, $p=0.5835$). In contrast, agent response variance was positively associated with absolute component gap ($r=0.4635$, $p=0.0298$; $\rho=0.5280$, $p=0.0116$). This suggests that TUX is most informative on headers that induce differentiated agent responses, rather than on headers that merely produce broad human disagreement. For benchmark construction, this points toward selecting headers that elicit meaningful variation in agent-side interpretations while preserving semantic coherence.

\begin{figure*}[!t]
\centering
\includegraphics[width=0.72\textwidth]{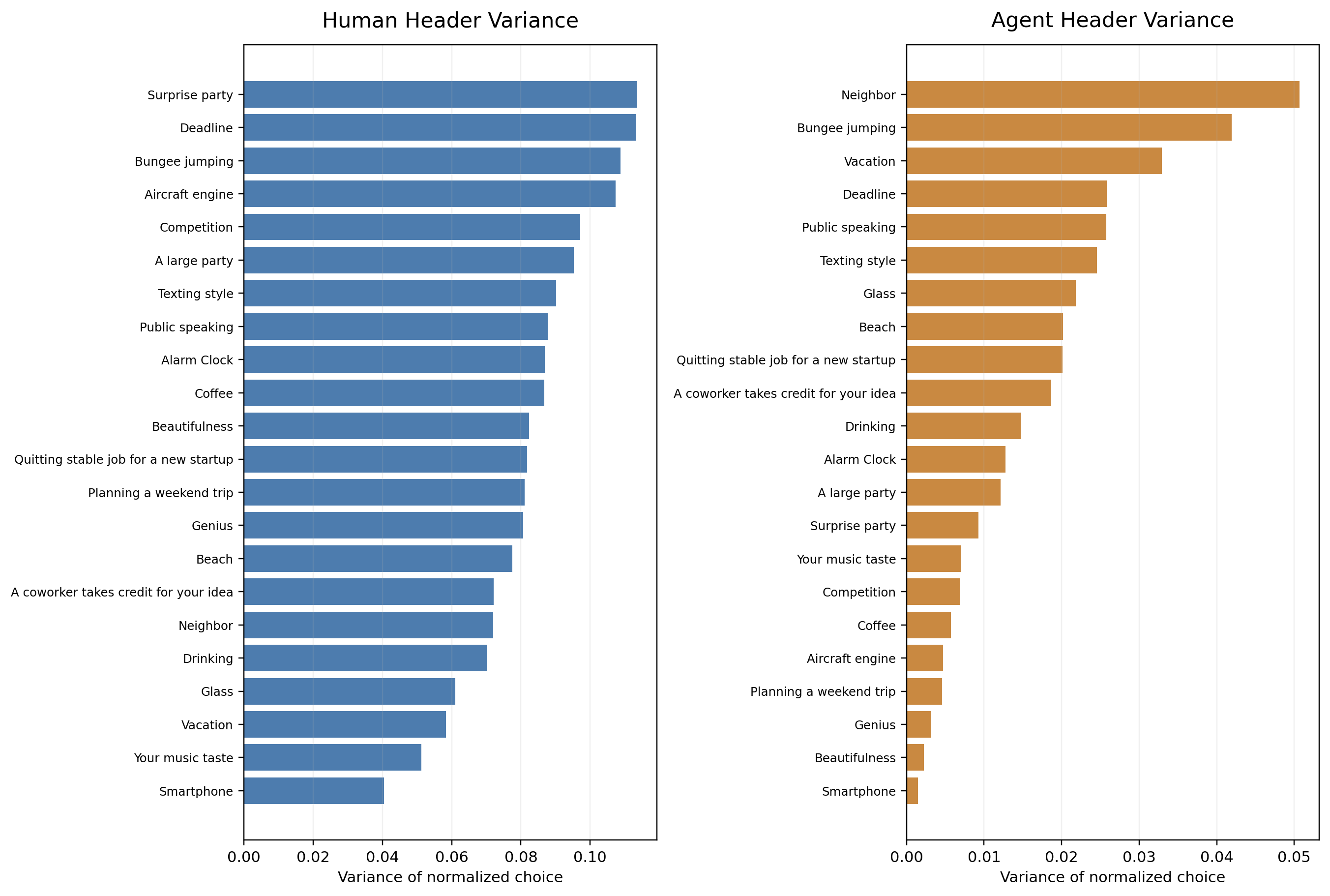}
\caption{Per-header response variance for humans and agents. Agent-side variance is positively associated with the absolute per-header matched-random TUX component gap, suggesting that headers with more differentiated agent responses carry more TUX-relevant signal.}
\label{fig:header_response_variance}
\end{figure*}

\subsection{Header-Specific Trait Diagnostics}
\label{app:header_specific_diagnostics}

We also examined whether some game headers are more informative than others for tacit-understanding benchmark construction. The results suggest substantial heterogeneity across headers. At the header level, trait explainability varied markedly: for example, under the descriptor-gap model, \textit{A large party} and \textit{Surprise party} reached $R^2=0.1609$ and $0.1625$, respectively, whereas \textit{Neighbor} and \textit{Beach} were much lower at $R^2=0.0165$ and $0.0262$. Adding aligned header-specific confidence often increased explainability substantially, with representative gains such as \textit{Public speaking} from $0.1190$ to $0.2276$, \textit{Alarm Clock} from $0.0697$ to $0.2026$, and \textit{Deadline} from $0.0541$ to $0.1881$.

Reverse-prediction analyses further showed that broader header sets carry meaningfully richer trait signal than the shared core-6 subset. In the human--agent pair setting, using all 22 header distances outperformed the \edit{core-6 subset} for every primary trait (e.g., extraversion: $R^2=0.1834$ vs.\ $0.1379$; agreeableness: $0.1111$ vs.\ $0.0415$). The same pattern held in the human population setting (e.g., extraversion: $0.4260$ vs.\ $0.3641$; rational ability: $0.2046$ vs.\ $0.0818$) and in the agent population setting (e.g., agreeableness: $0.2873$ vs.\ $0.0528$; extraversion: $0.2844$ vs.\ $0.1230$). Taken together, these diagnostics suggest that header informativeness is far from uniform, and that future tacit-understanding benchmarks may benefit from richer prompt sets rather than relying only on a small universally shared subset.

\section{Additional Human-Side Analyses}
\label{app:human_side}

\subsection{Human GDMS Pre/Post Shift}
\label{app:gdms_prepost}

As a human-only appendix analysis, we examined whether participants' self-reported decision-making style shifted from the pre-task survey to the post-task reflection items. This analysis asks whether completing spectrum-placement task changed how participants described their decision process. For each GDMS dimension, we computed the paired change score as $\Delta_i = \mathrm{post}_i - \mathrm{pre}_i$ and tested whether the mean change differed from zero using paired $t$-tests, with Wilcoxon signed-rank tests as a nonparametric robustness check.

\begin{figure*}[!t]
\centering
\includegraphics[width=0.62\textwidth]{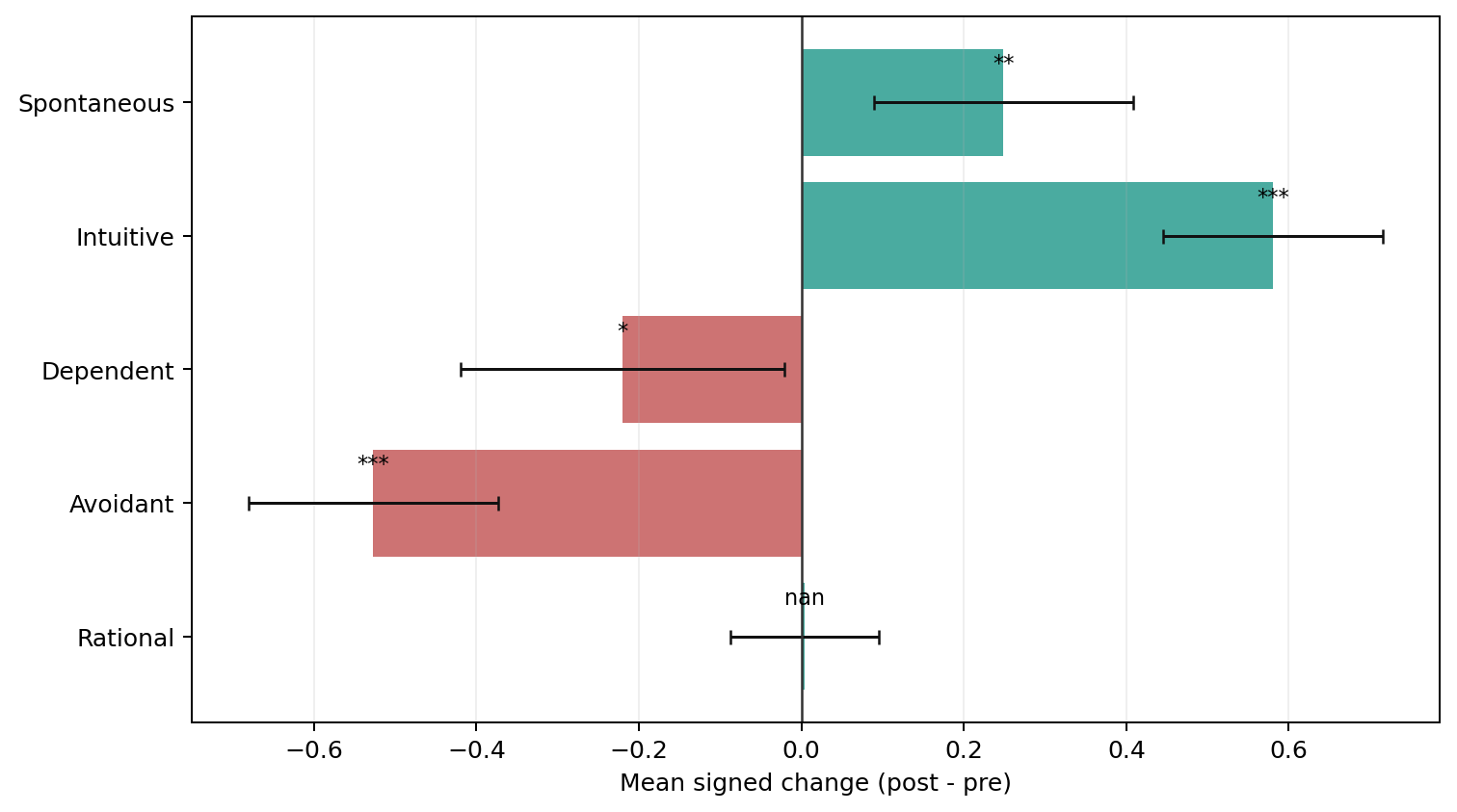}
\caption{Mean pre/post GDMS-style change among human participants. Positive values indicate higher post-task scores than pre-task scores; error bars show 95\% confidence intervals.}
\label{fig:gdms_prepost_change}
\end{figure*}

The clearest shifts were a decrease in avoidant style and increases in intuitive and spontaneous style (Figure~\ref{fig:gdms_prepost_change}). Avoidant style decreased substantially from pre-task to post-task ($\Delta=-0.5270$, 95\% CI $[-0.6807,-0.3733]$, $t=-6.7543$, $p<0.001$, Cohen's $d_z=-0.4351$). Intuitive style increased by $\Delta=0.5809$ (95\% CI $[0.4457,0.7162]$, $t=8.4067$, $p<0.001$, $d_z=0.5450$), and spontaneous style increased by $\Delta=0.2490$ (95\% CI $[0.0891,0.4088]$, $t=3.0673$, $p=0.0024$, $d_z=0.1976$). Rational style was effectively unchanged ($\Delta=0.0041$, $p=0.9291$), while the dependent-style proxy decreased modestly ($\Delta=-0.2199$, $p=0.0306$, $d_z=-0.1401$).

These results suggest that spectrum-placement task may prompt participants to describe their behavior as less avoidant and more intuition-driven after completing the task. However, the post-task items are not uniformly literal repeats of the pre-task GDMS scale. In particular, the avoidant item is proxy-comparable through time pressure, and the dependent-style item is based on reliance on task cues rather than interpersonal dependence. We therefore interpret these shifts as descriptive evidence about participants' perceived task strategy, not as evidence about human--agent alignment or TUX validity.

\end{document}